\documentclass[twocolumn]{article}

\usepackage{graphicx}
\usepackage[justification=justified]{subcaption}
\usepackage{amsmath, amssymb, amsthm}
\usepackage{accents}			

\usepackage[utf8]{inputenc}
\usepackage{acro}

\usepackage{pgfplots}
\usepgfplotslibrary{fillbetween}

\usepackage{algorithm,algorithmic}
\usepackage{enumerate}
\usepackage{xcolor}

\newtheoremstyle{bfplain}{}{}{\itshape}{}{\bfseries}{.}{ }{\thmname{#1}\thmnumber{ #2}\thmnote{ (#3)}}
\newtheoremstyle{bfdefinition}{}{}{}{}{\bfseries}{.}{ }{\thmname{#1}\thmnumber{ #2}\thmnote{ (#3)}}
\newtheoremstyle{itdefinition}{}{}{}{}{\itshape}{.}{ }{\thmname{#1}\thmnumber{ #2}\thmnote{ (#3)}}

\DeclareMathOperator*{\argmin}{arg\,min}

\DeclareMathOperator*{\st}{s.t.}

\newcommand{\satisfies}{\vDash}
\renewcommand{\and}{\wedge}

\NewDocumentCommand{\until}{oo}{
	\IfNoValueTF{#1}{U}{U_{[#1,#2]}}
}
\NewDocumentCommand{\eventually}{oo}{
	\IfNoValueTF{#1}{F}{F_{[#1,#2]}}
}
\NewDocumentCommand{\always}{oo}{
	\IfNoValueTF{#1}{G}{G_{[#1,#2]}}
}

\newcommand{\lbar}[1]{\underaccent{\bar}{#1}}
\newcommand{\diff}{\mathrm{d}}

\newcommand{\tp}{^{\textsc{T}}}

\newcommand{\inv}{^{-1}}
\renewcommand{\iff}{\Leftrightarrow}
\newcommand{\dotproduct}[2]{\langle#1, #2\rangle}

\newcommand{\bmat}[1]{\begin{bmatrix} #1 \end{bmatrix}}
\newcommand{\subnorm}[1]{_{#1}}
\NewDocumentCommand{\norm}{som}{
	\IfBooleanTF {#1}{\IfNoValueTF{#2} {\| #3 \|}{\| #3 \|\subnorm{#2}}}
	{\IfNoValueTF{#2} {	\left\| #3 \right\|}{\left\| #3 \right\|\subnorm{#2}}}
}
\NewDocumentCommand{\inReal}{soo}{
	\IfBooleanTF {#1}{}{\in}
	\mathbb{R}
	\IfNoValueTF{#2}{}{\IfNoValueTF{#3}{^{#2}}{^{#2 \times #3}}}	
}
\NewDocumentCommand{\inComplex}{soo}{
	\IfBooleanTF {#1}{}{\in}
	\mathbb{C}
	\IfNoValueTF{#2}{}{\IfNoValueTF{#3}{^{#2}}{^{#2 \times #3}}}	
}

\newcommand{\submat}[1]{_{\scriptstyle{#1}}}
\newcommand{\subvec}[1]{_{\scriptstyle{#1}}}
\newcommand{\supmat}[1]{^{\scriptstyle{#1}}}
\newcommand{\supvec}[1]{^{\scriptstyle{#1}}}

\newcommand{\mat}[1]{\mathbf{#1}}
\renewcommand{\vec}[1]{\boldsymbol{#1}}

\newcommand{\defvec}[2] {
	\DeclareDocumentCommand{#1}{s d<> d[] d''} {
		\IfBooleanTF {##1}
		{\IfNoValueTF{##2}{#2}{##2{#2}}}
		{\IfNoValueTF{##2}{\vec{#2}}{##2{\vec{#2}}}}		
		\IfNoValueTF{##3}{}{\subvec{##3}}
		\IfNoValueTF{##4}{}{\supvec{##4}}
	}
}

\newcommand{\defmat}[2] {
	\DeclareDocumentCommand{#1}{s d<> d[] d''} {
		\IfBooleanTF {##1}
		{\IfNoValueTF{##2}{#2}{[##2{#2}]}}
		{\IfNoValueTF{##2}{\mat{#2}}{##2{\mat{#2}}}}			
		\IfNoValueTF{##3}{}{\submat{##3}}
		\IfNoValueTF{##4}{}{\supmat{##4}}
	}
}
\defmat{\A}{A}
\defmat{\B}{B}
\defmat{\C}{C}
\defmat{\D}{D}
\defmat{\E}{E}
\defmat{\F}{F}
\defmat{\G}{G}
\defmat{\H}{H}
\defmat{\I}{I}
\defmat{\J}{J}
\defmat{\K}{K}
\defmat{\L}{L}
\defmat{\M}{M}
\defmat{\P}{P}
\defmat{\Q}{Q}
\defmat{\R}{R}
\defmat{\S}{S}
\defmat{\T}{T}
\defmat{\U}{U}
\defmat{\V}{V}
\defmat{\W}{W}
\defmat{\X}{X}
\defmat{\Y}{Y}
\defmat{\Z}{Z}
\defmat{\NMAT}{0}
\defmat{\SIG}{\Sigma}
\defmat{\LAM}{\Lambda}

\defvec{\ones}{1}
\defvec{\a}{a}
\defvec{\b}{b}
\defvec{\c}{c}
\defvec{\d}{d}
\defvec{\e}{e}
\defvec{\f}{f}
\defvec{\g}{g}
\defvec{\h}{h}
\defvec{\k}{k}
\defvec{\l}{l}
\defvec{\m}{m}
\defvec{\n}{n}
\defvec{\q}{q}
\defvec{\p}{p}
\defvec{\r}{r}
\defvec{\s}{s}
\defvec{\t}{t}
\defvec{\u}{u}
\defvec{\v}{v}
\defvec{\w}{w}
\defvec{\x}{x}
\defvec{\y}{y}
\defvec{\z}{z}
\defvec{\nvec}{0}
\defvec{\lam}{\lambda}
\defvec{\nuvec}{\nu}
\defvec{\pivec}{\pi}
\defvec{\phivec}{\phi}
\defvec{\rhovec}{\rho}
\defvec{\sigvec}{\sigma}
\defvec{\thetavec}{\theta}
\defvec{\alphavec}{\alpha}
\defvec{\gammavec}{\gamma}
\defvec{\Gammavec}{\Gamma}

\theoremstyle{bfdefinition}
\newtheorem{theorem}{Theorem}
\newtheorem{lemma}{Lemma}
\newtheorem{definition}{Definition}
\newtheorem{problem}{Problem}
\newtheorem{assumption}{Assumption}

\theoremstyle{itdefinition}
\newtheorem{remark}{Remark}

\pgfplotsset{compat=newest}
\pgfplotsset{every tick label/.append style={font=\small}}

\colorlet{darkgreen}{green!75!black}
\colorlet{lightblue}{blue!66!white}
\colorlet{lightred}{red!77!white}
\DeclareAcronym{TL}{
	short = TL,
	long = temporal logic
}
\DeclareAcronym{LTL}{
	short = LTL,
	long = linear temporal logic
}
\DeclareAcronym{TLTL}{
	short = TLTL,
	long = truncated linear temporal logic
}
\DeclareAcronym{STL}{
	short = STL,
	long = signal temporal logic
}
\DeclareAcronym{PPC}{
	short = PPC,
	long = prescribed performance control
}
\DeclareAcronym{PI2}{
	short = {PI$^2$},
	long = policy improvement with path integrals
}
\DeclareAcronym{TLPS}{
	short = TLPS,
	long = temporal logic policy search
}
\DeclareAcronym{ReLU}{
	short = ReLU,
	long = rectified linear unit
}
\DeclareAcronym{RL}{
	short = RL,
	long = reinforcement learning
}
\DeclareAcronym{MPC}{
	short = MPC,
	long = model predictive control
}
\DeclareAcronym{MDP}{
	short = MDP,
	long = Markov decision process
}
\DeclareAcronym{HJB}{
	short = HJB,
	long = Hamilton-Jacobi-Bellman
}

\usepackage[scaled=.95]{helvet}

\usepackage{geometry}
\usepackage{natbib}

\renewcommand\section{\@startsection{section}{1}{\z@}%
	{-3.5ex \@plus -1ex \@minus -.2ex}%
	{2.3ex \@plus.2ex}%
	{\normalfont\large\bfseries}}
\renewcommand\subsection{\@startsection{subsection}{1}{\z@}%
	{-3.5ex \@plus -1ex \@minus -.2ex}%
	{2.3ex \@plus.2ex}%
	{\normalfont\large\itshape}}

\title{\LARGE \bf Guided Policy Improvement for Satisfying STL Tasks using Funnel Adaptation}

\date{}

\author{Peter Varnai and Dimos V. Dimarogonas\footnotemark[1]}

\begin{document}

	\twocolumn[
	\begin{@twocolumnfalse}
		\maketitle
		\vspace{0.3cm}
		\begin{abstract}
			\noindent
			\textit{We introduce a sampling-based learning method for solving optimal control problems involving task satisfaction constraints for systems with partially known dynamics. The control problems are defined by a cost to be minimized and a task to be satisfied, given in the language of \ac{STL}. The complex nature of possible tasks generally makes them difficult to satisfy through random exploration, which limits the practical feasibility of the learning algorithm. Recent work has shown, however, that using a controller to guide the learning process by leveraging available knowledge of system dynamics to aid task satisfaction is greatly beneficial for improving the sample efficiency of the method. Motivated by these findings, this work introduces a controller derivation framework which naturally leads to computationally efficient controllers capable of offering such guidance during the learning process. The derived controllers aim to satisfy a set of so-called} robustness specifications \textit{or} funnels \textit{that are imposed on the temporal evolutions of the atomic propositions composing the \ac{STL} task. Ideally, these specifications are prescribed in a way such that their satisfaction would lead to satisfaction of the \ac{STL} task. In practice, however, such ideal funnels are not necessarily known} a priori\textit{, and the guidance the controller offers depends on their estimates. This issue is hereby addressed by introducing an adaptation scheme for automatically updating the funnels during the learning procedure, thus diminishing the role of their initial, user-specified values. The effectiveness of the resulting learning algorithm is demonstrated by two simulation case studies.}
\newline
\newline
		\end{abstract}
	
	\newline
	\newline
	\newline
	\end{@twocolumnfalse}
	]

	
	


	\section{Introduction} \label{section_introduction}

\Acp{TL} have been shown to be a powerful tool for expressing complex tasks and desired behaviors in a diverse range of applications in robotics. Recent examples include areas in control such as mobile service robots \citep{lacerda2019probabilistic}, generating collective swarm behaviors \citep{moarref2017decentralized}, task and motion planning for robotic systems \citep{saha2018task}, and hybrid systems \citep{filippidis2016control}. This paper examines controller design for nonlinear systems subject to tasks specified by \acf{STL}, a type of \acl{TL} originally introduced in the context of formal verification and monitoring the evolution of system signals \citep{maler2004monitoring}. \ac{STL} task specifications are composed of logical predicates depending on real-valued functions of the system states, and allow expressing explicit timing requirements to describe the desired system behavior.

Ensuring the satisfaction of \ac{STL} task specifications through proper controllers has been the topic of much research. Potential approaches for controller synthesis range from \ac{MPC} \citep{sadraddini2015robust, cho2018learning} to barrier function- \citep{lindemann2019control} and \acf{PPC}-based methods \citep{lindemann2017prescribed}. Generally, these methods require knowledge of the system dynamics, and there is a trade-off between their computational complexity and the range of system dynamics and \ac{STL} task fragments they are applicable to.

In addition to traditional control methods, \ac{RL} techniques have also gained attention in the temporal logics community, such as for learning to satisfy \ac{LTL} tasks \citep{sadigh2014learning}. An advantage of using an \ac{STL} task description for robotic systems is that \ac{STL} is equipped with various continuous robustness measures that express the degree of task satisfaction for system trajectories \citep{donze2010robust}; unlike in \ac{LTL}, where progress is only measured in discrete steps within a so-called Büchi automaton. Therefore, \ac{STL} robustness measures inherently serve as more descriptive rewards to be maximized for achieving task satisfaction through learning. Besides Q-learning approaches \citep{aksaray2016q}, policy search methods are being intensively studied as an alternative to deep learning methods \citep{sigaud2019policy, mania2018simple}, and have emerged in temporal logics as well \citep{fu2017sampling}. Inspired by successful results in their application for solving \ac{STL} tasks \citep{li2018policy}, this work also focuses on a particular type of policy search method named \ac{PI2} \citep{theodorou2010generalized}.

\footnotetext{$^{1}$Both authors are with the Division of Decision and Control Systems, School of Electrical Engineering and Computer Science, KTH Royal Institute of Technology, 114 28 Stockholm, Sweden. \\
	
\noindent
Corresponding author: \\
Peter Varnai \\ 
\noindent
Email: {\tt\small varnai@kth.se}}%

\newpage
More specifically, we consider the control problem for a system that is subject to satisfying an \ac{STL} task specification while minimizing a cost of interest, such as the expended input effort.  Preliminary results have appeared in \cite{varnai2019gradient, varnai2019prescribed}. In \cite{varnai2019prescribed}, it was shown that employing analytical control laws to guide the \ac{PI2} learning algorithm for solving \ac{STL} tasks can lead to significant improvements in terms of convergence rate, algorithm robustness, and general performance of the learning procedure. In \cite{varnai2019gradient}, the focus was placed on deriving computationally efficient controllers that guarantee satisfaction of simple subtasks and whose combination still serves as an effective guidance law for more complicated ones. Ultimately, however, the paper concluded that striving for such theoretical guarantees makes the individual controllers too aggressive and restrictive, diminishing the quality of the guidance they offer when taking their combination. 

This work builds upon the previous observations to provide significant improvements for the guided \ac{PI2} algorithm. Towards this end, a penalty-based controller derivation framework is introduced for devising guiding controllers in a structured manner. The aim for such controllers is to satisfy a set of user-defined so-called \textit{robustness specifications} or \textit{funnels}, which define how the robustness measures associated with atomic propositions composing the \ac{STL} task should evolve in time in order to achieve task satisfaction. The controllers are derived by minimizing a penalty term associated with the violation of these robustness specifications in a greedy fashion for computational efficiency. In case of unicycle-like dynamics, the resulting controllers are shown to yield the same improved guidance previously argued for heuristically in \cite{varnai2019gradient}. A further aspect to consider is if robustness specifications that are relevant for task satisfaction are difficult for the user to formulate in advance. To tackle this issue, we present a funnel adaptation scheme which automatically updates their initial estimates during the \ac{PI2} algorithm iterations in order to diminish the algorithm's reliance on them. The resulting adaptive \ac{PI2} algorithm is shown to yield superior performance, both in terms of achieved results and robustness of the algorithm.

The remainder of this paper is organized as follows. Section \ref{section:background} presents necessary background regarding \ac{STL} and \ac{PI2}, followed by a formal problem definition in Section \ref{section:problem}. Sections \ref{section:gradient_framework} and \ref{section:penalty_framework} discuss various methods of deriving base controllers in order to guide learning in the \ac{PI2} algorithm. Section \ref{section:adaptation} presents \ac{PI2}, tailored for the purpose of solving optimal control problems with \ac{STL} task constraints, along with the funnel adaptation scheme proposed for improving its performance. Finally, a case study is presented in Section \ref{section:case_study}and the paper is concluded in Section \ref{section:conclusions}.

	\section{Preliminaries} \label{section:background}

\subsection{\Acf{STL}}

\ac{STL} is a form of temporal logic defined over continuous-time signals (\cite{maler2004monitoring}). The \textit{predicates} $\mu$, which compose a \textit{task specification} $\phi$, are non-temporal atomic elements that can take on either true($\top$) or false($\bot$) values. These are determined according to a corresponding function $h^{\mu}(\x):\inReal*[n] \rightarrow \inReal*$ as follows; $\mu := \top$ if $h^{\mu}(\x) \geq 0$, and $\mu := \bot$ if $h^{\mu}(\x) < 0$. The predicates are then recursively combined using both Boolean logical and temporal operators to define more complex expressions:
\begin{equation} \label{eq:stlDef}
\phi := \top \ |\  \mu \ |\ \neg \phi \ |\ \phi_1 \and \phi_2 \ |\ \phi_1 \until[a][b]\phi_2,
\end{equation}
where the symbol $|$ separates the different possible recursive definitions. The time bounds of the \textit{until} operator $\until[a][b]$ satisfy $a,b \in [0,\infty)$ with $a \le b$. A signal $\x(t)$ is said to satisfy an \ac{STL} expression $\phi$ at time $t$, written as $(\x, t) \satisfies \phi$, by the following semantics (\cite{maler2004monitoring}):
\begin{align*}
&(\x, t) \satisfies \mu &&\iff h^{\mu}(\x(t)) \ge 0 \\
&(\x, t) \satisfies \neg\phi &&\iff \neg((\x, t) \satisfies \phi) \\
&(\x, t) \satisfies \phi_1 \and \phi_2 &&\iff (\x, t) \satisfies \phi_1 \and (\x, t) \satisfies \phi_2 \\
&(\x, t) \satisfies  \phi_1 \until[a][b]\phi_2 &&\iff \exists t_1 \in [t+a, t+b] \ : \ (\x, t_1) \satisfies \phi_2\\
& && \quad \ \  \mathrm{and}\ (\x, t_2) \satisfies \phi_1 \  \forall t_2 \in [t, t_1]. 
\end{align*}
Other commonly used and expressive temporal operators include the notion of \textit{eventually} ($\eventually$) and \textit{always} ($\always$), which are defined as $\eventually[a][b]\phi = \top \until[a][b]\phi$ and $\always[a][b]\phi = \neg \eventually[a][b] \neg \phi$.

For its role in a learning algorithm, an important aspect of \ac{STL} is that it is equipped with various robustness metrics which give an indication of how well a task specification is satisfied. Recently, much attention has been given to such metrics, with new definitions arising depending on the application domain; for examples, see \cite{akazaki2015time} or \cite{mehdipour2019arithmetic}. In this work, we employ the so-called \textit{spatial robustness} metric as defined in \cite{donze2010robust} and evaluated as follows for the formulas considered herein:
\begin{align*}
\rho^\mu(\x, t) &:= \rho^{\mu}(\x(t)) =  h^{\mu}(\x(t)) \\
\rho^{\neg \phi}(\x, t) &= -\rho^{\phi}(\x,t) \\
\rho^{\phi_1 \and \phi_2}(\x, t) &= \min\left(\rho^{\phi_1}(\x, t),\rho^{\phi_2}(\x, t)\right) \\
\rho^{\eventually[a][b]\phi}(\x, t)  &= \max_{t' \in [t+a,t+b]}\rho^{\phi}(\x,t') \\
\rho^{\always[a][b]\phi}(\x, t)  &= \min_{t' \in [t+a,t+b]}\rho^{\phi}(\x,t').
\end{align*}
The definitions are such that an \ac{STL} expression $\phi$ is satisfied at time $t$ if and only if the corresponding robustness metric $\rho^{\phi}(\x, t) \ge 0$, i.e., if the metric is non-negative. The exact value of the metric, however, gives a further indication of how well the task is satisfied or how much it is violated; thus, it contains more information than a true or false answer.

Depending on the \ac{STL} expression, it might take a finite or infinite amount of time to determine its truthfulness (and robustness). For example, the truth value of the formula $\eventually[0][2] (\norm{\x} \le 1)$ can be evaluated within $2s$, while $\always[0][\infty] (\norm{\x} \le 1)$ could turn out to be false at any $t > 0$. We refer to the maximum amount of time it takes to evaluate the truthfulness of an \ac{STL} expression as its \textit{time horizon}.

\subsection{\Acf{PI2}} \label{background:pi2}

The \ac{PI2} algorithm is a form of an evolutionary reinforcement learning method by which a system can find a solution to an optimization problem autonomously. The algorithm was originally introduced in \cite{theodorou2010generalized}, and has been further improved by incorporating covariance matrix adaptation in \cite{stulp2012path}. In the following, we outline the main steps of this latter variant.

Given a dynamical system $\x<\dot> = \f(\x, \u)$ and an objective $J(\cdot)$ to minimize, \ac{PI2} seeks to find an optimal policy in the following general form:
\begin{equation} \label{eq:policyForm}
\pi_{\theta}(\x, t) = \u<\hat>(\x, t) + \k_{\theta}(t).
\end{equation} 
The control action at each time step is thus composed of a \textit{base law} $\u<\hat>(\x, t)$ and a \textit{feedforward term} $\k_{\theta}(t)$. Both terms can be parameterized; however, for computational reasons the search is generally conducted only with respect to the feedforward term in order to find the optimal policy. In this work, we consider a parameterization that allows degrees of freedom for every time step $t$ within a time horizon $T$, i.e., $\theta = [\theta_0,\ \dots, \theta_T]$ using which the feedforward term is expressed as $\k_{\theta}(t) = \sum_{\tau=0}^{t} \theta_{\tau}$.

\ac{PI2} is initialized with an estimate $\theta^{(0)}$ of the optimal parameter vector and a probabilistic distribution around it which controls the exploration of the algorithm. A simple and common choice is to define a Gaussian distribution $\mathcal{N}(\theta^{(0)}_{t}, \C[t]^{(0)})$ for each time step $t$ with mean $\theta^{(0)}_{t}$ and a chosen covariance $\C[t]^{(0)}$. The main steps at iteration $k$ of the algorithm are then summarized as follows.
\begin{itemize}
	\item A set of controller parameters $\tilde{\theta}_{(i)} = [\tilde{\theta}_{0,i},\ \dots, \tilde{\theta}_{T,i}]$, $i = 1\dots N$, are evaluated from the current solution estimate by sampling $\tilde{\theta}_{t,i}$ from the Gaussian distribution $\mathcal{N}(\theta^{(k-1)}_t, \C^{(k-1)}_t)$ for each time step $t$.
	\item A cost $J_i$ is computed for each sampled parameter $\tilde{\theta}_{(i)}$ from the objective $J(\cdot)$ after running the system under the control law $\pi_{\tilde{\theta}_{(i)}}(\x, t) = \u<\hat>(\x, t) + \k_{\tilde{\theta}_{(i)}}(t)$ defined by $\tilde{\theta}_{(i)}$.
	\item The costs serve to assign weights $w_i$ (favoring the more optimal trajectories) to each sampled parameter, e.g.:
	\begin{equation} \label{eq:solution_weights}
	\w*[i] = \frac{e^{-\frac{1}{\eta} J_i}}{\sum_{j=1}^{N}e^{-\frac{1}{\eta} J_j}}.
	\end{equation}
	A higher value of the so-called \textit{temperature parameter} $\eta > 0$ will enhance the differences in the obtained costs and thus aim more towards minimizing the objective $J(\cdot)$.
	\item The weights are used to update the current estimate of the solution, as well as the exploration distributions for each time step $t$ by applying covariance matrix adaptation using weighted averaging \citep{stulp2012path}:	
	\begin{subequations}
		\begin{align} \label{eq:PI2paramUpdate} 
		\hspace{-3mm}\theta_{t}^{(k)} &= \sum_{i=1}^{N} \w*[i] \tilde{\theta}_{t,i}, \\ \vspace{-2mm}
		\hspace{-3mm}\C[t]'(k)' &=  \C[t,\min] + \sum_{i=1}^{N} \w*[i] (\tilde{\theta}_{t,i} - \theta_{t}^{(k)}) (\tilde{\theta}_{t,i} - \theta_{t}^{(k)})\tp.
		\end{align}
	\end{subequations}
	The term $\C[t,\min]$ enforces a minimal amount of exploration in subsequent iterations.
\end{itemize}

\newpage
These steps are iterated a given number of $K$ times, or until subsequent solution estimates differ marginally from one another (e.g. $\norm{\theta^{(k)}-\theta^{(k-1)}}/\norm{\theta^{(k)}} \le \epsilon$), implying convergence of the algorithm.

There are multiple factors which determine the convergence rate of \ac{PI2}, such as the form of the objective function or the feedforward parameterization and its update scheme. The base law $\u<\hat>(\x, t)$ can also contribute significantly to the achieved performance by guiding exploration, and its choice for the purpose of satisfying \ac{STL} task specifications constitutes the main topic of this paper.
	
	\section{Problem formulation} \label{section:problem}

\subsection{System and task description}

We consider nonlinear systems of the form:
\begin{equation} \label{eq:systemDynamics}
	\x<\dot> = \f(\x) + \g(\x)\u + \w, \qquad \x(0) = \x[0],
\end{equation}
where $\x \inReal[n]$ is the system state starting from an initial state $\x[0] \inReal[n]$, $\u \inReal[m]$ is the system input, and $\w \in \mathcal{W} \subset \inReal*[n]$ is bounded process noise. Given a time horizon $T$, a trajectory $\tau$ of the system is defined by the state and input signals $\x(t)$ and $\u(t)$ during the time $t \in [0, T]$. The following assumptions are imposed in order to guarantee existence and uniqueness of solutions.
\begin{assumption} \label{assumptions:dynamics}
	The functions $\f(\x)$ and $\g(\x)$ in the system dynamics (\ref{eq:systemDynamics}) are locally Lipschitz continuous and the noise $\w(t)$ is piecewise continuous in time.
\end{assumption}

The goal is to control the system such that the state trajectory $\x(t)$ satisfies a given \ac{STL} task $\phi$, i.e., $(\x, 0) \satisfies \phi$. The task is defined by $i = 1,\dots,M$ atomic predicates $\mu_i$ in a recursive manner similarly to (\ref{eq:stlDef}) as:
\begin{equation} \label{eq:stlSpecification}
	\phi := \top \ | \ \mu_i \ | \ \neg \phi \ | \ \phi_1 \and \phi_2 \ | \ \phi_1 \until[a][b]\phi_2,
\end{equation}
and satisfies the following properties.
\begin{assumption} \label{assumptions:task}\
	\begin{enumerate}[(i)]
		\item The task $\phi$ has a time horizon of finite length $T$, and
		\item the robustness metrics associated to the atomic predicates $\mu_{i}$ are such that each $\rho^{\mu_i}(\x)$ and its gradient $\frac{\partial \rho^{\mu_i}(\x)}{\partial \x}$ are locally Lipschitz continuous.
	\end{enumerate}
\end{assumption}
\subsection{Problem definition}

In this work, we are interested in controlling partially unknown dynamical systems in an optimal manner with respect to a given cost function while also satisfying an \ac{STL} task specification. A formal mathematical problem statement is given as follows.

\begin{problem}[Optimal \ac{STL} controller synthesis] \label{problemFormulation}
	Consider the dynamical system (\ref{eq:systemDynamics}) subject to an \ac{STL} task specification $\phi$ of the form (\ref{eq:stlSpecification}), and assume that Assumptions \ref{assumptions:dynamics} and \ref{assumptions:task} hold. Devise a control policy $\pi(\x, t) : \inReal*[n] \times \inReal* \rightarrow \inReal*[m]$ that drives the system to satisfy the task with a given minimal robustness $\rho^{\phi}(\x, 0) \ge \rho_{\min} \ge 0$ while minimizing a target cost $C(\tau)$ of the generated system trajectory using only knowledge of the term $\g(\x)$ in the system dynamics.
\end{problem}

The robustness $\rho^{\phi}(\x, 0)$ is a function of the trajectory $\tau$ and will be referred to as $\rho^{\phi}$ for simplicity. Note that the term $\f(\x)$ and the noise $\w$ are considered unknown, which motivates the use of a learning-based solution approach.

\subsection{Solution approach}

The outlined Problem \ref{problemFormulation} is a constrained optimal control problem in a continuous time, space, and input domain. The constraint is given by the minimal robustness $\rho_{\min}$ by which the system has to satisfy the \ac{STL} task $\phi$. Due to the incomplete knowledge of system dynamics, traditional optimal control methods such as \ac{MPC} are not applicable, and we turn our attention to learning-based approaches.

In particular, inspired by successful applications of policy improvement methods for \ac{STL} task satisfaction in \cite{li2018policy}, and by growing interest towards the class of such evolutionary methods in general \citep{mania2018simple, sigaud2019policy}, here we also pursue this direction and base our solution on the \ac{PI2} algorithm review in Section \ref{background:pi2}. Compared to other \acl{RL} methods such as Q-learning, this choice is also motivated by:
\begin{enumerate}[(i)]
	\item the need for an episode-based learning method, because the robustness measure of an \ac{STL} formula can generally only be evaluated at the end of its time horizon, and
	\item the ability of probability-weighted sampling methods such as \ac{PI2} to cope well with possibly high-dimensional search spaces.
\end{enumerate}

A crucial component of \ac{PI2} is the base control law $\u<\hat>(\x, t)$ in the control policy (\ref{eq:policyForm}), because it serves to guide the learning process and thus greatly impacts its performance. Following the \acf{PPC}-based approach introduced in \cite{lindemann2017prescribed}, we derive a class of controllers that aim to enforce the \ac{STL} task constraint, thus allowing the algorithm to explore in a more directed manner towards minimizing the target cost $C(\tau)$. These controllers rely on properly controlling the temporal evolution of (non-temporal) atomic propositions in order to satisfy the temporal \ac{STL} task at hand. Mathematically, this is expressed in the form of so-called \textit{robustness specifications} $\rho^{\mu_i}(\x(t)) \ge \gamma_i(t)$ laid on each $i = 1, \dots, M$ atomic proposition $\mu_i$, which are to be enforced given the curves $\gamma_i(t)$ as parameters. The base control law should rely on partial system information and be computationally efficient to evaluate. The latter is important for the performance of the learning process, which relies on simulating a large number of sample trajectories.

In the following two sections, we present two frameworks for deriving base controllers for a set of dynamical systems. The first has been introduced in our previous work \cite{varnai2019gradient}, where it was useful for deriving theoretical guarantees for satisfying robustness specifications. The second is a penalty-based approach which (heuristically) yields controllers more suitable for the purpose of guiding exploration. Afterwards, we present the \ac{PI2} algorithm adapted for solving Problem \ref{problemFormulation}, and a strategy for updating the $\gamma_i(t)$ parameterizations of the derived base laws in order for their guidance to remain relevant throughout algorithm iterations.

	\section{Gradient-based \ac{STL} control framework} \label{section:gradient_framework}

We begin by reviewing a gradient-based controller derivation framework originally inspired by the \ac{PPC} and barrier function methods of \cite{lindemann2017prescribed} and \cite{lindemann2019control}. The framework can be used to derive base laws which guarantee satisfaction of a \textit{single} robustness specification for a specific class of dynamical systems, which limits its practical use. Next, we hereby further improve these previous results by proposing an effective method for combining elementary control actions from such individual robustness specifications. This allows the computation of a controller that provides good guidance towards jointly satisfying a set of multiple robustness specifications. 

\subsection{Individual robustness specifications}

Consider the system (\ref{eq:systemDynamics}) subject to a single robustness specification $\rho^{\mu}(\x(t)) \ge \gamma(t)$ that is to be enforced by a proper controller. It is enough to control the system when this inequality constraint is close to being violated, motivating the following definition:
\begin{definition}[Region of interest] \label{def:roi}
	Let $\varGamma(t)$ be a smooth curve for which $\varGamma(t) \ge \gamma(t) + \epsilon$ for all $t \ge 0$ and some $\epsilon > 0$. The \textit{region of interest} $\mathcal{X}(t)$ at time $t$ is defined as:
	\begin{equation}
	\mathcal{X}(t) := \left\{\x \inReal[n] : \gamma(t) \le \rho^{\mu}(\x) \le \varGamma(t)\right\}.
	\end{equation}
	The upper and lower boundaries of this region are denoted by the two sets $\bar{\mathcal{X}}(t) := \left\{\x \inReal[n] : \rho^{\mu}(\x) = \varGamma(t)\right\}$ and $\lbar{\mathcal{X}}(t) := \left\{\x \inReal[n] : \rho^{\mu}(\x) = \gamma(t)\right\}$. We also introduce the \textit{uncontrolled region} $\mathcal{A}(t) := \left\{\x \inReal[n] : \rho^{\mu}(\x) > \varGamma(t)\right\}$.
\end{definition}

Next, we define a notion of locally satisfying robustness specifications, which will be useful for examining local, gradient-based control laws.
\begin{definition}[Local robustness satisfaction]
	Let the system (\ref{eq:systemDynamics}) be controlled by $\u = \u(\x,t)$. This control law is said to locally satisfy the robustness specification $\rho^{\mu}(\x(t)) \ge \gamma(t)$ in a domain $\mathcal{D} \subseteq \inReal*[n]$ if, for any $t'$ and $\x(t') \in \mathcal{D}$ such that $\rho^{\mu}(\x(t')) \ge \gamma(t')$, there exists a time $\delta > 0$ for which $\rho^{\mu}(\x(t)) \ge \gamma(t)$ holds during $t \in [t', t'+\delta]$.
\end{definition}
\noindent
We now present a class of controllers which achieve local robustness satisfaction of an individual robustness specification for systems of the form (\ref{eq:systemDynamics}) under certain controllability assumptions.

The available control input $\u(t)$ appears in the time derivative of the robustness metric $\rho^{\mu}(t)$:
\begin{equation} \label{eq:robustnessDerivative}
\dot{\rho}^{\mu}(\x) = \dfrac{\partial \rho^{\mu}(\x)}{\partial \x} \left(\f(\x) + \g(\x)\u + \w\right).
\end{equation}
To ease notation, define the coefficient of $\u$ in this expression as:
\begin{equation} \label{eq:vDef}
\v(\x)\tp := \dfrac{\partial \rho^{\mu}(\x)}{\partial \x} \g(\x).
\end{equation}
Furthermore, let us consider systems and task specifications which allow direct control over the robustness derivative (\ref{eq:robustnessDerivative}), i.e., where the following assumption holds:
\begin{assumption} \label{assumption:controllability}
	For the term $\v(\x)$ in (\ref{eq:vDef}), it holds that:
	\begin{equation} \label{eq:controllability}
	\v(\x) \ne \nvec, \quad \forall \x : \exists t\ \st \ \x \in \mathcal{X}(t).
	\end{equation}
\end{assumption}
\noindent
This controllability assumption may already be violated if $\g(\x)$ is not full row rank or if $\rho^{\mu}(\x)$ has a singularity in the region of interest. Nevertheless, we note that it is equivalent to the standard relative degree one condition made with \ac{PPC} and control barrier functions, such as in \cite{xu2015robustness}.

The main result of this section can now be stated as follows.
\begin{theorem} \label{theorem:localSatisfaction}
	Let Assumptions \ref{assumptions:dynamics}, \ref{assumptions:task} (ii), and \ref{assumption:controllability} hold. Define
	\begin{equation} \label{eq:baseControlFamily}
	\u(\x,t) := \begin{cases}
	\nvec \qquad &\text{if } \x \in \mathcal{A}(t), \\
	\kappa(\x, t) \dfrac{K}{\norm[2]{\v(\x)}^{2} + \varDelta}\v(\x) &\text{if } \x \notin \mathcal{A}(t),
	\end{cases}
	\end{equation}
	where the coefficient $\kappa(\x, t) \ge 0$ is continuous in $t$, locally Lipschitz in $\x$, and satisfies (i) $\kappa(\x, t) \ge \dot{\gamma}(t) + B(\x)$ with $B(\x) \ge -\frac{\partial \rho^{\mu}(\x)}{\partial \x} f(\x) + \max_{\w}\norm[2]{\frac{\partial \rho^{\mu}(\x)}{\partial \x} \w}$ for all $\x \in \lbar{\mathcal{X}}(t)$ and (ii) $\kappa(\x, t) = 0$ for all $\x \in \bar{\mathcal{X}}(t)$. Then, with a proper choice of the additional parameters $K \ge 1$ and $\varDelta \ge 0$, this control law achieves local robustness satisfaction of the specification $\rho^{\mu}(\x(t)) \ge \gamma(t)$ for the system (\ref{eq:systemDynamics}) in the entire domain $\inReal*[n]$.
\end{theorem}

The class of controllers defined by (\ref{eq:baseControlFamily}) can be considered as a set including two extremes, namely whether the inequality (i) for $\kappa(\x, t)$ is satisfied as an exact equality or independently of $\x$ using $\kappa(\x, t) \rightarrow \infty$ as $\x \rightarrow \lbar{\mathcal{X}}(t)$. These two extremes roughly correspond to the barrier function or \ac{PPC} approaches of \cite{lindemann2019control} and \cite{lindemann2017prescribed}, respectively. However, the derived class of controllers also allows intermediate cases, which we have advocated is better for the performance in guiding learning algorithms in \cite{varnai2019prescribed}. In particular, the form of functions used in our simulation studies are joint linear and sigmoid curves of the form:
\begin{equation} \label{eq:functionS}
\kappa(\x,t) = m\xi + \dfrac{\vartheta_1}{1 + e^{\vartheta_2(\xi - 1)}}
\end{equation}
where $\xi = \frac{\Gamma(t) - \rho^{\mu}(\x(t))}{\Gamma(t) - \gamma(t)}$ is a normalized metric of how close the system is to violating the robustness specification $\rho^{\mu}(\x(t)) \ge \gamma(t)$. This function takes the value $m + \vartheta_1$ at the robustness specification boundary $\xi = 1$ (when $\x(t) \in \lbar{\mathcal{X}}(t)$) $\xi = 1$, and the steepness near this transition is dictated by $\vartheta_2$.

\subsection{Multiple robustness specifications} \label{section_combinatinons}

In this section, we propose a method for combining elementary control actions $\u[i]$ calculated from $i = 1, \dots, M$ individual robustness specifications using the derived controller (\ref{eq:baseControlFamily}). The goal is to obtain a single control action which aims to achieve local robustness satisfaction of all specifications $\rho^{\mu_i}(\x(t)) \ge \gamma_{i}(t)$ in a computationally efficient manner. For simplicity, assume that the $M$ robustness specifications at time $t$ are all actively being controlled in the sense that $\x(t) \notin \mathcal{A}_i(t)$ for all $i$. The notation from the previous section is maintained, with subscript $i$ being used to refer to the $(i)$-th specification.

The simple method proposed in our original approach \cite{varnai2019gradient} for combining elementary control actions is the weighted sum $\u = \sum_i \alpha_{i} \u[i]$. Higher weights are assigned to control inputs stemming from robustness specifications closer to violation; a sample choice is $\alpha_i = \frac{\Gamma_i(t) - \rho^{\mu_i}(\x(t))}{\Gamma_i(t) - \gamma_i(t)}$. Substituting in the derived form (\ref{eq:baseControlFamily}), this \textit{simple combination controller} can be expressed as:
\begin{equation} \label{eq:simpleCombination}
\u(\x, t) = \sum_{i=1}^{M} \alpha_i \kappa_i(\x,t) \dfrac{K_i}{\norm[2]{\v[i](\x)}^2 + \Delta_i} \v[i](\x).
\end{equation}

In order to introduce and motivate a more practical method for controller combination, consider the following. At any moment in time, the contribution of the preferred control action $\u[i]$ to the robustness derivative $\dot{\rho}^{\mu_i}(\x)$ of the $(i)$-th robustness measure is given by:
\begin{equation}
\dot{\rho}^{\mu_i}_{\u[i]}(\x) = \dfrac{\partial \rho^{\mu_i}(\x)}{\partial \x} g(\x) \cdot \u[i](\x, t) = \dotproduct{\v[i](\x)}{\u[i](\x,t)}.
\end{equation}
On the other hand, when applying a combined input $\u$ to the system, the actual contribution of the input to the derivative will be:
\begin{equation}
\dot{\rho}^{\mu_i}_{\u}(\x) = \dfrac{\partial \rho^{\mu_i}(\x)}{\partial \x} g(\x) \cdot \u(\x, t) = \dotproduct{\v[i](\x)}{\u(\x, t)}
\end{equation}
The difference between these two terms can be expressed as:
\begin{equation} \label{eq:rhoDotDiff}
\Delta \dot{\rho}^{\mu_i}_{\u} = \dot{\rho}^{\mu_i}_{\u} - \dot{\rho}^{\mu_i}_{\u[i]} = \dotproduct{\v[i](\x)}{\u(\x,t) - \u[i](\x, t)}
\end{equation}
and can generally be expected to take on a nonzero value, because the elementary control actions will most likely differ from the combined control action. However, an approximate least-violating locally task satisfying control action can be chosen as the solution to an optimization problem aiming to minimize these differences:
\begin{equation} \label{eq:optProblem}
\u(\x, t) := \argmin_{\u} \sum_{i = 1}^M \dfrac{1}{2}\alpha_i \left(\Delta \dot{\rho}^{\mu_i}_{\u} \right)^2.
\end{equation}
Here, the weights $\alpha_i$ are again chosen to give the most violating specification the most weight, as in the simple combination case. The term \textit{approximate} stems from the fact that both positive and negative $\Delta \dot{\rho}^{\mu_i}_{\u}$ derivative differences are penalized, whereas this is actually only necessarily for the latter in order for the combined control action to avoid violating all the robustness specifications. On the other hand, the least squares problem (\ref{eq:optProblem}) offers a computationally efficient solution, which can be derived as follows.

Substituting in (\ref{eq:rhoDotDiff}) for $\Delta \dot{\rho}^{\mu_i}_{\u}$ in the optimization problem, the term to minimize becomes:
\begin{equation}
\sum_{i = 1}^M \dfrac{1}{2}\alpha_i \dotproduct{\v[i]}{\u - \u[i]} \cdot \dotproduct{\v[i]}{\u- \u[i]}.
\end{equation}
Setting the gradient with respect to the input $\u$ to zero, the optimal solution must satisfy:
\begin{equation}
\sum_{i = 1}^{M} \alpha_i \v[i]\v[i]\tp \u - \sum_{i=1}^{M} \alpha_i \v[i]\v[i]\tp \u[i] = \nvec.
\end{equation}
The identity $\v[i] \v[i]\tp \u[i] = \v[i]\tp \v[i] \u[i]$, which can be readily derived from the form (\ref{eq:baseControlFamily}) of the individual control actions $\u[i]$ for the case $\x(t) \notin \mathcal{A}_i(t)$, allows this to be rewritten as:
\begin{equation}
\sum_{i = 1}^{M} \alpha_i \v[i]\v[i]\tp \u - \sum_{i=1}^{M} (\alpha_i \v[i]\tp\v[i]) \u[i] = \nvec.
\end{equation}
The solution can thus be found by solving the linear matrix equation:
\begin{equation}
\left(\sum_{i = 1}^{M} \alpha_i \v[i]\v[i]\tp\right) \u = \sum_{i=1}^{M} (\alpha_i \v[i]\tp\v[i]) \u[i].
\end{equation}
Substituting in the individual control actions $\u[i]$ from the form (\ref{eq:baseControlFamily}), an alternative form can be obtained as:
\begin{equation} \label{eq:combinedEquation1}
\left(\sum_{i = 1}^{M} \alpha_i \v[i]\v[i]\tp\right) \u = \sum_{i=1}^{M} \alpha_i \kappa_i(\x,t) \dfrac{K_i \norm{\v[i]}^{2}}{\norm{\v[i]}^{2} + \Delta_i} \v[i].
\end{equation}
Note, however, that the $\Delta_i \ge 0$ terms were originally introduced to allow avoiding numerical issues near singular configurations of $\v_i \rightarrow \nvec$, with high enough $K_i \ge 1$ values still allowing individual robustness satisfaction guarantees. In the matrix equation (\ref{eq:combinedEquation1}), numerical issues are potentially caused by an ill-conditioned coefficient matrix. Thus, by setting each $\Delta_i = 0$ and $K_i = 1$ and instead augmenting the optimization problem (\ref{eq:optProblem}) with a regularization term $\u\tp \Delta \u$ on the sought after input in its entirety, the equation becomes:
\begin{equation}
\left(\sum_{i = 1}^{M} \alpha_i \v[i]\v[i]\tp + \Delta \I\right) \u = \sum_{i=1}^{M} \alpha_i \kappa_i(\x,t) \v[i].
\end{equation}
This is guaranteed to have a unique solution for $\Delta > 0$ due to the coefficient matrix becoming positive definite on the left hand side, and thus an \textit{improved combination controller} can be obtained as:
\begin{equation} \label{eq:improvedCombination}
\u = \left(\sum_{i = 1}^{M} \alpha_i \v[i]\v[i]\tp + \Delta \I\right)\inv \left(\sum_{i=1}^{M} \alpha_i \kappa_i(\x,t) \v[i]\right).
\end{equation}

\begin{figure}[b]
	\centering
	\begin{subfigure}{0.54\linewidth}
		\centering
		\begin{tikzpicture}
		\begin{axis}[width=0.8\linewidth, height=\linewidth,grid=both,legend style={font=\scriptsize},legend pos=outer north east, axis equal,xmin=1,ymin=1,ymax=4,xmax=3.6,xlabel=$x$,ylabel=$y$,axis x line=bottom, axis y line = left,ytick={1,...,4},xtick={1,...,3},ylabel near ticks,ylabel style={rotate=-90},minor x tick num=1, minor y tick num = 1,grid style={line width=.2pt, draw=gray!50}, enlarge y limits={abs=0.4}, major grid style={line width=.4pt,draw=gray!80}]
		
		\filldraw[fill=black,draw=black] (1.2,0.7) circle (.03);
		\filldraw[fill=black,draw=black] (2.0,0.8) circle (.03);
		\filldraw[fill=black,draw=black] (3.0,0.8) circle (.03);
		\filldraw[fill=red,draw=red,fill opacity=0.5] (2.5,2.5) circle (1);
		\filldraw[fill=green!66!black,draw=green!66!black,fill opacity=0.5] (2.0,4.2) circle (0.1);
		\filldraw[fill=green!66!black,draw=green!66!black,fill opacity=0.5] (3.0,4.2) circle (0.1);
		
		\addplot [line width=2pt, lightblue, mark=none, opacity=0.8] table[x=x1,y=x2] {data/complex/sampleSimpleController.dat};
		\addlegendentry{Ground1}
		\addplot [line width=2pt, orange!80!black, mark=none, opacity=0.8] table[x=x3,y=x4] {data/complex/sampleSimpleController.dat};
		\addlegendentry{Ground2}
		\addplot [line width=2pt, red!50!blue!70!white, mark=none, opacity=0.8] table[x=x5,y=x6] {data/complex/sampleSimpleController.dat};
		\addlegendentry{Drone}
		
		\end{axis}
		\end{tikzpicture}
		\caption{$\rho^{\phi} = -2.53$}
	\end{subfigure}
	\begin{subfigure}{0.44\linewidth}
		\centering
		\begin{tikzpicture}
		\begin{axis}[width=\linewidth, height=1.227\linewidth,grid=both,axis equal,xmin=1,ymin=1,ymax=4,xmax=3.6,xlabel=$x$,ylabel=$y$,axis x line=bottom, axis y line = left,ytick={1,...,4},xtick={1,...,3},ylabel near ticks,ylabel style={rotate=-90},minor x tick num=1, minor y tick num = 1,grid style={line width=.2pt, draw=gray!50}, enlarge y limits={abs=0.4}, major grid style={line width=.4pt,draw=gray!80}]
		
		\filldraw[fill=black,draw=black] (1.2,0.7) circle (.03);
		\filldraw[fill=black,draw=black] (2.0,0.8) circle (.03);
		\filldraw[fill=black,draw=black] (3.0,0.8) circle (.03);
		\filldraw[fill=red,draw=red,fill opacity=0.5] (2.5,2.5) circle (1);
		\filldraw[fill=green!66!black,draw=green!66!black,fill opacity=0.5] (2.0,4.2) circle (0.1);
		\filldraw[fill=green!66!black,draw=green!66!black,fill opacity=0.5] (3.0,4.2) circle (0.1);
		
		\addplot [line width=2pt, lightblue, mark=none, opacity=0.8] table[x=x1,y=x2] {data/complex/sampleImprovedController.dat};
		\addplot [line width=2pt, orange!80!black, mark=none, opacity=0.8] table[x=x3,y=x4] {data/complex/sampleImprovedController.dat};
		\addplot [line width=2pt, red!50!blue!70!white, mark=none, opacity=0.8] table[x=x5,y=x6] {data/complex/sampleImprovedController.dat};

		\end{axis}
		\end{tikzpicture}
		\caption{$\rho^{\phi} = -0.59$}
	\end{subfigure}
	\caption{Performance of the (a) simple and (b) improved combination controllers (\ref{eq:combinedEquation1}) and (\ref{eq:improvedCombination}) for satisfying the \ac{STL} task $\phi$ in the complex scenario described in Appendix \ref{appendix:complexScenario}. The latter controller achieves much better results, and thus provides better guidance towards task satisfaction, as implied by the obtained higher robustness metric $\rho^{\phi}$.}
	\label{fig:controllerCombination}
\end{figure}
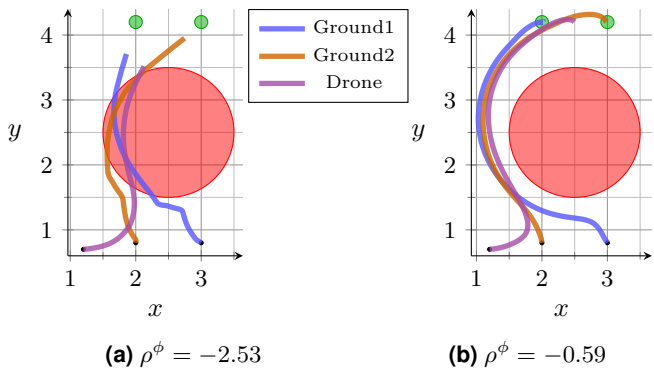

Figure \ref{fig:controllerCombination} shows a sample improvement this controller achieves over the simple weighted average combination (\ref{eq:simpleCombination}) for the complex navigation scenario described in Appendix \ref{appendix:complexScenario}. The new controller is able to provide much better guidance towards satisfying a set of $M=7$ robustness specifications and thus a given \ac{STL} task $\phi$. This is seen from the greatly improved robustness metric $\rho^{\phi}$ of the obtained system trajectory, although task satisfaction with $\rho^{\phi} \ge 0$ is not achieved in either case.

It is important to emphasize, however, that the derived controllers are \textit{not} required to guarantee satisfaction of the robustness specifications. Their role is important more in the sense of providing good guidance in a computationally efficient manner when serving as a base law for guiding learning during the \ac{PI2} algorithm. We argue that this fact should be considered in the controller design phase, especially regarding systems with more complex dynamics, for which inexpensive controllers with theoretical guarantees are not necessarily possible to derive.

	\section{Penalty-based \ac{STL} control framework} \label{section:penalty_framework}

In this section, we re-derive the previously obtained controllers from an alternative point of view inspired by a penalty-based learning formulation for satisfying robustness specifications. In particular, we impose a penalty for violating the specifications, and show that minimizing the resulting regularized optimization problem in a greedy manner yields a class of controllers of the form (\ref{eq:baseControlFamily}). We advocate that this penalty-based framework leads more naturally to the improved controller combination (\ref{eq:improvedCombination}), and is also readily extendable to systems with more complex dynamics. The discussion is again separated first for the case of individual robustness specifications, then for the combination of multiple ones.

\subsection{Individual robustness specifications}

Consider again the system (\ref{eq:systemDynamics}), restated for convenience as:
\begin{equation} \label{eq:systemDynamics_restated2}
\x<\dot> = f(\x) + g(\x)\u + \w.
\end{equation}
Imposing a single robustness specification $\rho^{\mu}(\x(t)) \ge \gamma(t)$ on the system, we define a penalty function $P(\rho^{\mu}(\x), t) : \mathbb{R} \times \mathbb{R} \rightarrow \mathbb{R}$ to penalize the violation of this specification; i.e., $P$ increases as $\rho^{\mu}(\x)$ decreases and nears the boundary of the inequality. Intuitively, with large enough penalties, minimizing
\begin{equation}
	\int_{0}^{T} P(\rho^{\mu}(\x), t) \diff t
\end{equation}
for the problem's time horizon $T$ could then lead to satisfaction of the specification. Such a problem is in general difficult and computationally expensive to solve; the solution can be obtained, for example, using dynamic programming or learning-based methods. 

Aiming to derive a computationally efficient guiding law, we instead adopt a greedy strategy. At a time instance $t$, we attempt to decrease the integrand through the available input $\u$ by minimizing a cost $J(\u)$ composed of its time derivative $\dot{P}$ and an added regularization term in the form:
\begin{equation} \label{eq:singlePenaltyOptimization}
	\min_{\u} J(\u) = \min_{\u} \dot{P}(\rho^{\mu}(\x), t) + \dfrac{1}{2}\u\tp \R(\x)\u.
\end{equation}
The regularization term $\R(\x) > \NMAT$ serves to avoid actuator saturation and numerical issues regarding singular cases. The main result of this section is now summarized as follows.

\begin{theorem} \label{theorem:penaltyController}
	The solution to the optimization problem (\ref{eq:singlePenaltyOptimization}) takes the form:
	\begin{equation} \label{eq:generalPenaltyController}
		\u(\x, t) = -\dfrac{\partial P(\rho^{\mu}(\x), t)}{\partial \rho^{\mu}}\R(\x)\inv\v(\x).
	\end{equation}	
	Furthermore, with a regularization term defined by either of the matrices $\R{'}(\x) = K^{-1}(\norm[2]{\v(\x)}^2 + \Delta)\I$ or $\R{''}(\x) = K^{-1}(\v(\x) \v(\x)\tp + \Delta)\I$, this solution leads to the same form of the controller (\ref{eq:baseControlFamily}), namely:
	\begin{equation} \label{eq:matchedPenaltyController}
		\u(\x, t) = -\dfrac{\partial P(\rho^{\mu}(\x), t)}{\partial \rho^{\mu}}\dfrac{K}{\norm[2]{\v(\x)}^2 + \Delta}\v(\x).
	\end{equation}
\end{theorem}

Comparing (\ref{eq:matchedPenaltyController}) with (\ref{eq:baseControlFamily}), it is clear that the term $-\frac{\partial P(\rho^{\mu}(\x), t)}{\partial \rho^{\mu}}$ plays the same role as $\kappa(\x,t)$ previously. This implies that the theoretical guarantees for local robustness satisfaction are also recovered for a choice of penalty function $P(\rho^{\mu}(\x), t)$ which satisfies the same conditions as $\kappa(\x,t)$, i.e, whose partial derivative is non-negative, locally Lipschitz in $\x$, continuous in $t$, becomes $0$ at $\x \in \bar{\mathcal{X}}(t)$, and satisfies $-\frac{\partial P(\rho^\mu(\x), t)}{\partial \rho^\mu} \ge \dot{\gamma}(t) + B(\x)$ for all $\x \in \lbar{\mathcal{X}}(t)$.

The physical interpretation of the two regularization terms is as follows. In both cases, the regularization increase in case the input can more easily affect the system through $\v(\x)$, i.e., we penalize a large input if $\v(\x)$ has a large magnitude as well in order to avoid unnecessarily large control actions. With $\R{'}$, the penalization is the same across all entries of $\u$, whereas with $\R{''}$, the inputs impacting the system through $\v(\x)$ matter more substantially. We can expect this latter to be a more suitable choice for combining such elementary controllers for the case of multiple robustness specifications, since only the relevant input elements are being penalized.
\subsection{Multiple robustness specifications}

In this section, we show that the most natural extension for deriving a controller for a conjunction of robustness specifications using the penalty framework already yields the improved controller (\ref{eq:improvedCombination}) derived previously. 

\begin{theorem} \label{theorem:penaltyCombination}
	Following the form of the cost (\ref{eq:singlePenaltyOptimization}), for multiple robustness specifications $\rho^{\mu_i}(\x(t)) \ge \gamma_i(t)$, $i = 1,\dots, M$, define the optimization problem:
	\begin{equation} \label{eq:combinedPenaltyOptimization}
	\min_{\u} \sum_{i=1}^{M} \alpha_i \left(\dot{P}_i(\rho^{\mu_i}(\x), t) + \dfrac{1}{2}\u\tp \R[i](\x)\u\right),
	\end{equation}
	where $\alpha_i > 0$ are user-defined weights normalized to $\sum \alpha_i = 1$. The solution takes the same form as the improved combination controller (\ref{eq:improvedCombination}) with the input regularization choice $\R[i](\x) = \Delta \I + \v[i](\x)\v[i](\x)\tp$, i.e.:
	\begin{equation*}
		\u = \left(\sum_{i=1}^{M}\alpha_{i} \v[i] \v[i]\tp + \Delta \I\right) \inv \sum_{i=1}^{M} \left(-\alpha_{i}\dfrac{\partial P_i(\rho^{\mu_i}(\x), t)}{\partial \rho^{\mu_i}}\v[i]\right).
	\end{equation*}
\end{theorem}

We argue that the penalty framework is thus more suitable and natural for deriving practically efficient controllers than the gradient-based framework. Furthermore, the new framework potentially offers an additional advantage. Assuming trajectory costs in the form of instantaneous rewards $C(\tau) = \int_{0}^{T} c(\x) + c(\u) \diff t$, it may yield insight into how to incorporate the goal of minimizing such a cost into the controller derivation in order to provide further improved guidance for learning algorithms.
\subsection{Extension to nonholonomic dynamics} \label{section_dynamics}

The gradient-based framework can be used to derive controllers which guarantee local robustness satisfaction of individual robustness specifications for nonholonomic, unicycle-like systems under certain assumptions, as shown in \cite{varnai2019gradient}. However, therein it was argued that this guarantee can be detrimental to the guiding performance when combining such controllers for multiple specifications, and instead a heuristic controller was proposed. In the following, we show that the introduced penalty framework naturally leads to the same form of this latter suggested controller. 

The form of the dynamical system under consideration is 
\begin{equation} \label{eq:unicycleSystem}
\x<\dot> := \bmat{\x<\dot>[1] \\ \x<\dot>[2]} = \bmat{\f[1](\x[1]) \\ \f[2](\x)} + \bmat{\g[11](\x[2]) & \NMAT \\ \g[21](\x) & \g[22](\x)} \bmat{\u[1] \\ \u[2]} + \bmat{\w[1] \\ \w[2]},
\end{equation}
and the atomic predicate $\mu$ is assumed to have a robustness metric dependent only on $\x[1]$, i.e., $\rho^{\mu}(\x[1])$. For example, in the unicycle model with input velocity $v$ and steering velocity $\omega$:
\begin{equation}
\x<\dot> := \bmat{\dot{x} \\ \dot{y} \\ \dot{\theta}} = \bmat{\cos(\theta) & 0 \\ \sin(\theta) & 0 \\ 0 & 1} \bmat{v \\ \omega},
\end{equation}
and $\mu$ allows us to express propositions regarding the position of the unicycle.

In line with the penalty framework and considering a single robustness specification, we adapt a greedy strategy and minimize the time derivative of the imposed penalty term $P(\rho^{\mu}(\x[1]), t)$ for specification violation with respect to each regularized input, as in (\ref{eq:singlePenaltyOptimization}).

The first input $\u[1]$ has a direct effect on the derivative $\dot{P}$, and is determined using the previously derived controller (\ref{eq:penaltyController}), i.e., by solving:
\begin{equation} \label{eq:singlePenaltyOptimizationU1}
\min_{\u[1]} \dot{P}(\rho^{\mu}(\x[1]), t) + \dfrac{1}{2}\u[1]\tp \R[1](\x)\u[1],
\end{equation}  
to obtain:
\begin{equation} \label{eq:unicycleU1}
\u[1](\x, t) = -\dfrac{\partial P(\rho^{\mu}(\x[1]), t)}{\partial \rho^{\mu}} \R[1](\x)\inv \v[1](\x),
\end{equation}
where $\v[1](\x)\tp = \frac{\partial \rho^{\mu}(\x[1])}{\partial \x[1]} \g[11](\x[2])$ and the regularization term $\R[1](\x) = (\v[1](\x) \v[1](\x)\tp + \Delta_{1})\I$.

The input $\u[2]$ then aims to further decrease the penalty term by a natural extension of (\ref{eq:singlePenaltyOptimization}), formulating the optimization with respect to the second derivative of the penalty, in which $\u[2]$ appears.
\begin{theorem} \label{theorem:unicycleController}
	Consider the optimization problem
	\begin{equation} \label{eq:singlePenaltyOptimizationU2}
	\min_{\u[2]} \ddot{P}(\rho^{\mu}(\x[1]), t) + \dfrac{1}{2}\u[2]\tp \R[2](\x)\u[2],
	\end{equation}	
	where $\R[2](\x) = (\v[2](\x) \v[2](\x)\tp + \Delta_{2})\I$. The solution obtained by treating $\u[1]$ as a constant leads to the same form as the suggested heuristic controller in \cite{varnai2019gradient}, namely:
	\begin{equation} \label{eq:unicycleU2}
	\u[2](\x, t) = -\dfrac{\partial P(\rho^{\mu}(\x[1]), t)}{\partial \rho^{\mu}} \R[2](\x)\inv \v[2](\x, \u[1]),
	\end{equation}
	where the term $\v[2](\x, \u[1])\tp = \u[1]\tp \frac{\partial \v[1](\x)}{\partial \x[2]} \g[22](\x)$.
\end{theorem}

\begin{remark}
	Treating $\u[1] = \u[1](\x, t)$ as a constant is a conservative assumption. In case $\u[1]$ is computed from a combination of multiple specifications, rapid changes stemming from a specific specification are mitigated, and the assumption might be more reasonable. We emphasize, however, that the goal here is to obtain an empirically good and computationally efficient guidance controller, which motivates making such a simplifying assumption.
\end{remark}

\begin{remark}
	When considering the combination of multiple robustness specifications, the respective $\u[1]$ and $\u[2]$ inputs can be solved for in succession by forming the weighted sum of the individual penalty costs (\ref{eq:singlePenaltyOptimizationU1}) and (\ref{eq:singlePenaltyOptimizationU2}), in the same manner as previously done in (\ref{eq:combinedPenaltyOptimization}) for the cost (\ref{eq:singlePenaltyOptimization}). For the first input, the weights are chosen to give higher emphasis to constraints near violation, as discussed previously. However, for the second input, the effect of $\u[1]$ towards constraint satisfaction can also be taken into account. In this sense, if $\u[1]$ has already helped in achieving the desired increase in robustness for a given specification, less weight can be assigned to the corresponding second input, which would further aim to increase the robustness metric. The sample choice of weights for combining the elementary $\u[2]$ actions at a given time instance $t$ used in the case study is: 
	\begin{equation}
	\alpha_i = e^{-\nu \u[1]\tp\v[1,i](\x(t))}\cdot \dfrac{\Gamma_i(t) - \rho^{\mu_i}(\x[1](t))}{\Gamma_i(t) - \gamma_i(t)},
	\end{equation}
	The added exponential term diminishes the weight in case the direction of $\u[1]$ already aligns with the desired direction $\v[1,i]$ towards increasing the $(i)$-th robustness metric $\rho^{\mu_i}(\x[1])$. The decrease is controlled by a parameter $\nu > 0$; in the case study scenario it is set to $\nu = 5$.
\end{remark}
	
	\section{Policy improvement for \ac{STL} task satisfaction} \label{section:adaptation}

After analyzing derivations of computing base control laws for the purpose of guiding exploration in the \ac{PI2} algorithm, we now turn our attention to the algorithm itself. First, we review the form of \ac{PI2} proposed in our previous work \cite{varnai2019prescribed}, tailored for solving \ac{STL} tasks. Then, the solution algorithm is further extended by an adaptive strategy referred to as \textit{funnel adaptation}, which aims to keep the base law relevant as a guiding controller throughout the \ac{PI2} iterations.

\subsection{\ac{PI2} for solving \ac{STL} tasks}

In order to solve Problem \ref{problemFormulation}, the generic \ac{PI2} algorithm described in Section \ref{background:pi2} is adapted to aim at minimizing a trajectory cost $C(\tau)$ while enforcing the task satisfaction robustness constraint $\rho^{\phi} \ge \rho_{\min}$. This goal can be accomplished by a suitable choice of the cost function $J(\tau)$ that scores each trajectory sampled during the iterations of \ac{PI2}, namely, by adding a penalty term to the target cost $C(\tau)$ as\footnotemark:
\begin{equation} \label{eq:solution_objective}
J(\tau) := J^{\lambda}(\tau,\rho^{\phi}) = C(\tau) + P^{\lambda}(\rho^{\phi}).
\end{equation}
The penalty term $P^{\lambda}(\rho^{\phi})$ is parameterized by $\lambda > 0$ in a way such that $P^{\lambda}(\rho^{\phi} \ge \rho_{\min}) \rightarrow 0$ and $P^{\lambda}(\rho^{\phi} < \rho_{\min}) \rightarrow \infty$ as $\lambda \rightarrow \infty$. The task satisfaction constraint is then progressively enforced throughout the \ac{PI2} iterations by increasing $\lambda$.

From a numerical perspective, it is important to normalize the obtained trajectory costs in order to achieve faster convergence rates by properly discriminating between the $i = 1, \dots, N$ sampled trajectories. Denoting the cost (\ref{eq:solution_objective}) corresponding to the $(i)$-th sample by $J^{\lambda}_i$, the respective normalized cost is defined as:
\begin{equation} \label{eq:solution_normalization}
\bar{J}^{\lambda}_i := -h \eta \dfrac{J^{\lambda}_i - \min_j J^{\lambda}_j}{J^{\lambda}_{\epsilon} - \min_j J^{\lambda}_j}.
\end{equation}
In this expression, $\eta$ is the temperature parameter from the weight update equation (\ref{eq:solution_weights}) of \ac{PI2}, the parameter $h$ controls the range of the normalized values, and $J^{\lambda}_{\epsilon}$ is the value below which $\epsilon \%$ of all sampled $J^{\lambda}_i$ costs fall. The motivation behind this form of normalization, as opposed to the original method of using $\max_j J^{\lambda}_j$ in place of $J^{\lambda}_{\epsilon}$ in \cite{chebotar2017path}, can be explained as follows. Due to the high penalty values that trajectories can incur by violating the robustness constraint as $\lambda \rightarrow \infty$, some samples might be assigned extremely high costs. The normalization (\ref{eq:solution_normalization}) prevents these extreme cases from corrupting the discrimination between those sample trajectories that achieve lower costs, allowing \ac{PI2} to perform parameter updates in a more targeted manner towards minimizing the objective.
\footnotetext{Note that the robustness $\rho^{\phi}$ is also a function of the trajectory $\tau$, but its role is highlighted explicitly in the equations.}

The full solution algorithm to Problem \ref{problemFormulation}, augmented with a Nesterov acceleration scheme for improved convergence rate \citep{nesterov1983method} and the potential funnel adaptation procedure, is summarized as Algorithm \ref{alg:pi2solution} below.

\setlength{\skip\footins}{0mm}
\begin{algorithm}[h!]  
	\caption{Guided \ac{PI2} solution to Problem \ref{problemFormulation}} 
	\label{alg:pi2solution}
	\begin{algorithmic}[1]
		\REQUIRE Initial parameter estimates $\theta^{(0)}_t$, covariances $\C[t]'(0)'$, sample batch size $N$, iteration number $K$, penalty $\lambda$
		\STATE $\alpha^{(0)} := 1$, $\hat{\theta}_t^{(0)} := \theta_t^{(0)}$ for  $\forall t = 0, \dots, T$
		\STATE Perform funnel adaptation
		\FOR{$k = 1 \dots K$}
		\FOR{$i = 1 \dots N$}
		\STATE Sample policy parameters $\tilde{\theta}_{t,i}$ from the distribution $\mathcal{N}(\hat{\theta}_{t}^{(k-1)}, \C[t]'(k-1)')$ for each $t = 0, \dots, T$ to form the entire parameterization $\tilde{\theta}_{(i)} = [\tilde{\theta}_{0,i},\ \dots, \tilde{\theta}_{T,i}]$.
		\STATE Obtain the trajectory $\tau_i$ using the guided policy $\pi_{\tilde{\theta}_{(i)}}(\x, t) = \u<\hat>(\x, t) + \k_{\tilde{\theta}_{(i)}}(t)$ defined by $\tilde{\theta}_{(i)}$ \label{alg:policy}
		\ENDFOR
		\STATE Compute the normalized cost $\bar{J}^{\lambda}_{i}$ for each trajectory $\tau_{i}$ using (\ref{eq:solution_objective}) and (\ref{eq:solution_normalization})
		\STATE Compute weights $w_i = \frac{e^{-\frac{1}{\eta} \bar{J}^{\lambda}_{i}}}{\sum_j e^{-\frac{1}{\eta} \bar{J}^{\lambda}_{j}}}$
		\FOR{each time step $t = 0, \dots, T$}
		\STATE $\theta_{t}^{(k)} = \sum_{i=1}^{N} \w*[i] \tilde{\theta}_{t,i}$ 
		\STATE $\C[t]'(k)' =  \C[t,\text{min}] + \sum\limits_{i=1}^{N} \w*[i] (\tilde{\theta}_{t,i} - \theta_{t}^{(k)}) (\tilde{\theta}_{t,i} - \theta_{t}^{(k)})\tp$
		\STATE $\alpha^{(k)} = (1 + \sqrt{4 \left(\alpha^{(k-1)}\right)^2 + 1})/2$ \label{alg:Nesterov}
		\STATE $\hat{\theta}_{t}^{(k)} = \theta_{t}^{(k)} + (\alpha^{(k-1)} - 1)(\theta_{t}^{(k)} - \theta_{t}^{(k-1)}) / \alpha^{(k)}$ \label{alg:NesterovUpdate}
		\ENDFOR
		\STATE Increase penalty term $\lambda$, perform funnel adaptation
		\ENDFOR
		\RETURN found solution $\theta = \theta^{(K)} = [\theta_{0}^{(K)},\ \dots, \theta_{T}^{(K)}]$
	\end{algorithmic} 
\end{algorithm}
\setlength{\skip\footins}{1mm}
\subsection{Funnel adaptation}

The main purpose of the guiding control $\u<\hat>(\x,t)$ in \ac{PI2} is for it to aid enforcing the \ac{STL} task satisfaction constraint while the learning process searches for optimal control actions with respect to the target cost $C(\tau)$. The guidance is accomplished by attempting to enforce $M$ robustness specifications $\rho^{\mu_{i}}(\x(t)) \ge \gamma_{i}(t)$ using properly defined $\gamma_{i}(t)$ curves. This `proper' definition, however, is far from trivial, as enforcing the corresponding robustness specifications should ideally both:  
\begin{enumerate}[(i)]
	\item guarantee satisfaction of the \ac{STL} task $\phi$ with the required minimal robustness $\rho^{\phi} \ge \rho_{\min}$, and
	\item do so in a way such that the cost $C(\tau)$ is minimized.
\end{enumerate}
It is important to emphasize that these conditions do not need to hold for the defined robustness specifications, as they merely serve to parameterize the guiding controller and thus aid exploration during the \ac{PI2} algorithm. However, it can be expected that with a better estimate of the $\gamma_i(t)$ curves, the exploration becomes more targeted and the learning process converges more rapidly.

The guiding controllers (\ref{eq:simpleCombination}) or (\ref{eq:improvedCombination}) aiming to enforce the $i=1,\dots,M$ robustness specifications are explicit functions of time due to their dependency on the curves $\gamma_i(t)$ and $\Gamma_i(t)$ through the coefficient $\kappa(\x, t)$ (and possibly the weights $\alpha_i$). These curves define the robustness specifications and the regions of interest within which the controller aims to enforce them. Denoting the collection of these parameters by $\gammavec(t) = [\gamma_1(t), \dots, \gamma_M(t)]\tp$ and $\Gammavec(t) = [\Gamma_1(t), \dots, \Gamma_M(t)]\tp$, the general policy sought after by \ac{PI2} for solving Problem \ref{problemFormulation} may be written in the alternative form:
\begin{equation} \label{eq:policyFormSTL}
\pi(\x, t) = \u<\hat>(\x, \gammavec(t), \Gammavec(t)) + \k_{\theta}(t).
\end{equation}

Aiming towards gradually fulfilling the described ideal conditions (i) and (ii) regarding the purpose of the robustness specifications, we propose a technique termed \textit{funnel adaptation} to improve the quality of the guiding controller as the \ac{PI2} learning process progresses. More specifically, the guiding law parameters $\gammavec(t)$ and $\Gammavec(t)$ are continuously updated after each iteration such that they follow the evolution of the robustness metrics $\rho^{\mu_i}(\x(t))$ associated with the currently found most optimal trajectory. The adaptation is performed in a manner such that the overall control actions $\pi(\x, t) = \u<\hat>(\x, \gammavec(t), \Gammavec(t)) + \k_{\theta}(t)$ remain unaltered. In order to achieve this, the feedforward parameter $\theta$ is also updated and the difference between the current base control actions $\u<\hat>(\x, \gammavec(t), \Gammavec(t))$ and the updated base control actions $\u<\hat>(\x,  \gammavec{'}(t), \Gammavec{'}(t))$ is calculated and transferred to the new $\k_{\theta'}(t)$ for every time step $t$. The algorithm is summarized as Algorithm \ref{alg:funnelAdaptation} on the right. 

\begin{remark}
	The proposed form of funnel adaptation is only possible if the feedforward term parameterization $\theta$ allows for degrees of freedom during every time step in order to maintain the equality of the control actions before and after funnel adaptation, i.e., in order to solve for step \ref{step:thetaUpdate} of Algorithm \ref{alg:funnelAdaptation}. Further research into how the feedforward and feedback parameter updates could be done in an alternating fashion to allow lower-dimensional curve parameterizations is subject of ongoing work.
\end{remark}

Intuitively, every iteration of \ac{PI2} produces an improved solution towards minimizing $C(\tau)$ while satisfying the given \ac{STL} task, thus, adjusting the guiding parameterizations $\gammavec(t)$ and $\Gammavec(t)$ to this solution can be seen as a step towards satisfying the ideal robustness specification conditions. The exact method of choosing how to update these curves is a tuning procedure for the trade-off between using a base law which is aggressive enough to effectively keep enforcing relevant robustness specifications, but lenient enough to continue allowing exploration. Adapting $\gammavec(t)$ closer to the found solution's evolution aids the former, while placing $\Gammavec(t)$ closer aids the latter. We found a value for the goal transformed robustness measure (\ref{eq:xiDef}) of $\xi_t = 0.8$ to work well in this regard. This value is chosen such that with the linear-sigmoid transformation functions (\ref{eq:functionS}) used in our case studies, the new transformed robustness measure would be placed at the starting point of where the sigmoid curve increases and aims at enforcing the corresponding robustness specification by keeping $\xi < 1$. Note that there is an extra degree of freedom in satisfying (\ref{eq:xiDef}), which allows for additional design choices of, for example, keeping the width $\Gamma_i(t) - \gamma_i(t)$ of the control region $\mathcal{X}_i(t)$ constant or fixing the upper boundary $\Gamma_i(t)$.

The funnel adaptation algorithm is run at initialization and after every iteration of \ac{PI2}, as seen in steps 2 and 16 of Algorithm \ref{alg:pi2solution}. The benefits of employing the proposed funnel adaptation scheme are illustrated in the next section.

\begin{algorithm}
	\caption{Funnel adaptation}
	\label{alg:funnelAdaptation}
	\begin{algorithmic}[1]
		\REQUIRE Candidate optimal trajectory $\x(t)$ for time steps $t = 0 \dots T$; curves $\gamma_i(t)$, $\Gamma_i(t)$ and lower limits $\gamma_i^{\lim}(t)$ for $i = 1 \dots M$; target transformed robustness measure $\xi_t \in (0,1)$; averaging parameter $\beta$
		\FOR{$t = 0 \dots T$}
		\FOR{$i = 1 \dots M$}
		\STATE Calculate the robustness $\rho^{\mu_i}(\x(t))$ 
		\STATE Determine a new parameters $\tilde{\gamma}_i(t)$ and $\tilde{\Gamma}_i(t)$ such that the transformed robustness measure satisfies
		\begin{equation} \label{eq:xiDef}
		\dfrac{\tilde{\Gamma}_{i}(t) - \rho^{\mu_i}(\x(t))}{\tilde{\Gamma}_{i}(t) - \tilde{\gamma}_i(t)} = \xi_t
		\end{equation} 
		\STATE Obtain the new parameterization curves as $\gamma'_i(t) = \beta \tilde{\gamma}_i(t) + (1-\beta) \gamma_i(t)$ and $\Gamma'_i(t) = \beta \tilde{\Gamma}_i(t) + (1-\beta) \Gamma_i(t)$. The \textit{averaging parameter} $\beta$ controls the aggressiveness towards adapting the curves to their goal values and impacts the robustness of the adaptation procedure.\footnotemark
		\ENDFOR
		\STATE Find updated feed-forward parameters $\theta'$ such that the overall control action remains unchanged by satisfying $\u<\hat>(\x, \gammavec(t), \Gammavec(t)) + \k_{\theta}(t) = \u<\hat>(\x, \gammavec{'}(t), \Gammavec{'}(t)) + \k_{\theta'}(t)$ \label{step:thetaUpdate}
		\ENDFOR
		\STATE Set $\gammavec(t) = \gammavec{'}(t)$, $\Gammavec(t) = \Gammavec{'}(t)$, and $\theta = \theta'$
	\end{algorithmic}
\end{algorithm}
\footnotetext{The value $\gamma'_i(t)$ is also optionally clipped between the required $\rho_{\min}$ from above and the given limiting $\gamma^{\lim}_i(t)$ values from below. The former serves to prevent unnecessarily enforcing high robustness values, while the latter can be used to make sure that the new $\gamma'(t)$ does not deviate much from enforcing the desired temporal behavior.}

	\section{Case study} \label{section:case_study}

In the following, we present two case studies to illustrate the performance of the derived base control laws as guiding controllers for the developed \ac{PI2} algorithm. The first examines the benefits of funnel adaptation for both a single integrator and a unicycle system tasked with a simple navigation task. These benefits, along with the advantage of using the improved combination controller (\ref{eq:improvedCombination}) over (\ref{eq:simpleCombination}), are further illustrated in a more complex task scenario taken from \cite{varnai2019prescribed} for comparison.

\vspace{-2mm}
\subsection{Simple navigation task}\vspace{-1mm}

\begin{figure}[b]\vspace{-6mm}
	\centering
	\begin{tikzpicture}
	\begin{axis}[width=0.7\linewidth, height=0.7\linewidth,grid=both, axis equal,xmin=0.0,ymin=0,ymax=4.1,xmax=4.1,xlabel=$x$,ylabel=$y$,axis x line=bottom, axis y line = left,ytick={0,1,...,4},xtick={0,1,...,4},ylabel near ticks,ylabel style={rotate=-90},minor x tick num=1, minor y tick num = 1,grid style={line width=.2pt, draw=gray!50}, major grid style={line width=.4pt,draw=gray!80}]
	
	\filldraw[fill=black,draw=black] (3.5,0.3) circle (.03);
	\filldraw[fill=red,draw=red,fill opacity=0.5] (2.5,2) circle (1.2);
	\filldraw[fill=green!66!black,draw=green!66!black,fill opacity=0.5] (1.0,3.5) circle (0.2);
	
	\addplot [line width=2pt, lightblue, mark=none] table[x=x,y=y] {data/simpleSampleSolution.dat};
	
	\node at (2.5,2.15) {\color{red!50!black} \textbf{obstacle}};
	\node at (1,3.8) {\color{green!60!black} \textbf{goal}};
	\node at (3.8,0.35) {$\x[0]$};
	\node at (-0.3,2) {$y$};
	\node at (2.0,-0.4) {$x$};
	
	\end{axis}
	\end{tikzpicture}\vspace{-2mm}
	\caption{Depiction of the simple navigation task scenario along with a sample solution trajectory. The robot, starting at an initial $\x[0]$ position, must eventually reach and remain within the green goal region while always avoiding the red obstacle.}
	\label{fig:case1Sample}
\end{figure}
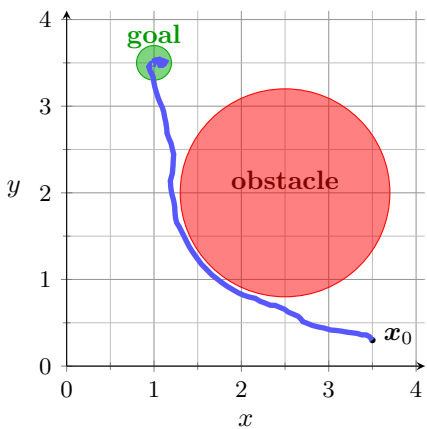

We first consider a simple navigation task in which a robot has to reach and remain within a goal region while passing by an obstacle during a time horizon of $T = 10$s. The scenario and a sample solution trajectory are shown in Figure \ref{fig:case1Sample}. The scenario is examined both in case the robot is modeled as a single integrator (omnidirectional vehicle) and as a unicycle. For both, the input constraints are chosen such that the speed of the robot is limited by $v \le 1$.

The advantage of funnel adaptation is illustrated by solving the scenario task while minimizing a family of different target cost functions of the form:
\begin{equation} \label{eq:case1Cost}
C(\tau) = \vartheta T^* + \int_{0}^{T} v(t)^2 \diff t.
\end{equation}
Here, $T^*$ is the first time instance after which the robot does not leave the goal region, $v(t)$ is its speed, and $\vartheta > 0$ defines the trade-off between reaching the goal quickly and minimizing the expended input effort to do so. The full scenario description, formal \ac{STL} task specification, guide controller parameters, and hyperparameters of the \ac{PI2} algorithm used to find solutions are given in Appendix \ref{appendix:simpleScenario}.

The navigation problem is instructive as it is simple enough to calculate the true optimal solutions corresponding to different target costs parameterized by $\vartheta$. It is clear that for any parameter value, the robot should (1) take the shortest possible path from $\x[0]$ to the goal region and then (2) immediately stop to further minimize its expended input energy. The robot's velocity $v$ during the first phase should be constant and is dictated by the trade-off defined by $\vartheta$.

Let the distance of the shortest path to the goal region be denoted by $D$; for the studied scenario, this path is composed of two straight lines connected by an arc and can be calculated to be $D \approx 4.37$. As the robot travels with a constant velocity $v$ during phase (1) of its motion, we have $T^* = D/v$. Substituting into the cost (\ref{eq:case1Cost}) and noting that the velocity must remain $0$ during the second motion phase (for $t > T^*$), $C(\tau)$ can be expressed as a function of the velocity $v$ as:\vspace{-1mm}
\begin{equation} \label{eq:optCost}\vspace{-1mm}
C(v) = \vartheta T^* + T^* v^2 = \vartheta \dfrac{D}{v} + \dfrac{D}{v} v^2 = \vartheta \dfrac{D}{v} + Dv.
\end{equation}
To find the optimal vehicle speed $v_{\text{opt}}$, note that we must have $v_{\text{opt}} \le 1$ due to the input constraints, and also $v_{\text{opt}} \ge D/T$ since the goal region must be reached within the time horizon $T$. Furthermore, as the cost $C(v)$ is a convex function of $v$ in case $v, \vartheta > 0$, the optimal velocity is the minimum of $C(v)$ projected onto the interval $[D/T, 1]$. Setting the derivative of (\ref{eq:optCost}) to zero, we have:\vspace{-1mm}
\begin{equation*}\vspace{-2mm}
\dfrac{\diff C(v)}{\diff v} = -\vartheta \dfrac{D}{v^2} + D = 0,
\end{equation*}
which yields $v_{\text{opt}} = \max(D/T, \min(1, \sqrt{\vartheta}))$ and allows the optimal cost to then be calculated using (\ref{eq:optCost}).

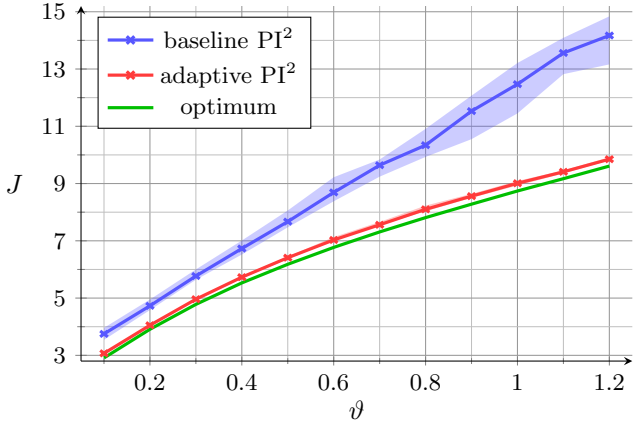
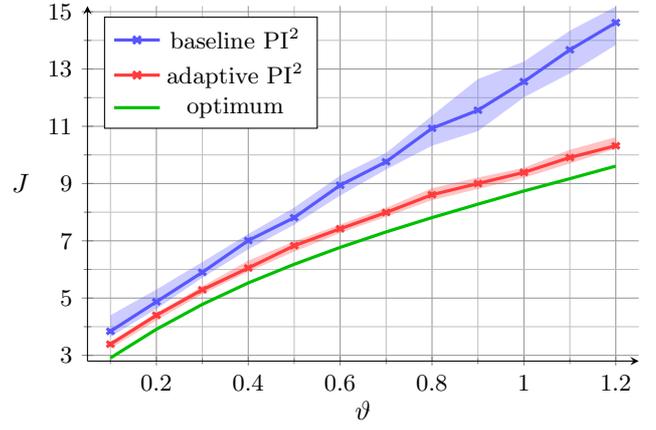
\begin{figure*}[t]
	\centering
	\begin{subfigure}[b]{0.48\linewidth}
		\centering
		\begin{tikzpicture}
		\begin{axis}[grid=both,width=\linewidth,xmin=0.1,ymin=3,ymax=15,xmax=1.2,xlabel=$\vartheta$,ylabel=$J$,legend pos=north west,axis x line=bottom, axis y line = left,ytick={3,5,...,15},ylabel near ticks,ylabel style={rotate=-90},minor x tick num=1, minor y tick num = 1, enlarge x limits={abs=0.05}, enlarge y limits={abs=0.2},grid style={line width=.2pt, draw=gray!50}, major grid style={line width=.4pt,draw=gray!80},yscale=0.78,legend style={font=\small}]
		\addplot [very thick, lightblue, mark=x] table[x=xi,y=JNF] {data/simple_w0.dat};
		\addlegendentry{baseline PI$^2$}
		\addplot [very thick, lightred, mark=x] table[x=xi,y=JF] {data/simple_w0.dat};
		\addlegendentry{adaptive PI$^2$}
		\addplot [very thick, darkgreen, mark=none] table[x=x,y=y] {data/simple_optimum.dat};
		\addlegendentry{optimum}
		
		\addplot [name path=upper,draw=none] table[x=xi,y expr=\thisrow{JNF_max}] {data/simple_w0.dat};
		\addplot [name path=lower,draw=none] table[x=xi,y expr=\thisrow{JNF_min}] {data/simple_w0.dat};
		\addplot [fill=lightblue, opacity=0.3] fill between[of=upper and lower];		
		\addplot [name path=upper,draw=none] table[x=xi,y expr=\thisrow{JF_max}] {data/simple_w0.dat};
		\addplot [name path=lower,draw=none] table[x=xi,y expr=\thisrow{JF_min}] {data/simple_w0.dat};
		\addplot [fill=lightred, opacity=0.3] fill between[of=upper and lower];		
		\end{axis}
		\end{tikzpicture}
		\caption{Without process noise}
	\end{subfigure}
	\begin{subfigure}[b]{0.48\linewidth}
		\centering
		\begin{tikzpicture}
		\begin{axis}[grid=both,width=\linewidth,xmin=0.1,ymin=3,ymax=15,xmax=1.2,xlabel=$\vartheta$,ylabel=$J$,legend pos=north west,axis x line=bottom, axis y line = left,ytick={3,5,...,15},ylabel near ticks,ylabel style={rotate=-90},minor x tick num=1, minor y tick num = 1, enlarge x limits={abs=0.05}, enlarge y limits={abs=0.2},grid style={line width=.2pt, draw=gray!50}, major grid style={line width=.4pt,draw=gray!80},yscale=0.78,legend style={font=\small}]
		\addplot [very thick, lightblue, mark=x] table[x=xi,y=JNF] {data/simple_w0.2.dat};
		\addlegendentry{baseline PI$^2$}
		\addplot [very thick, lightred, mark=x] table[x=xi,y=JF] {data/simple_w0.2.dat};
		\addlegendentry{adaptive PI$^2$}
		\addplot [very thick, darkgreen, mark=none] table[x=x,y=y] {data/simple_optimum.dat};
		\addlegendentry{optimum}
		
		\addplot [name path=upper,draw=none] table[x=xi,y expr=\thisrow{JNF_max}] {data/simple_w0.2.dat};
		\addplot [name path=lower,draw=none] table[x=xi,y expr=\thisrow{JNF_min}] {data/simple_w0.2.dat};
		\addplot [fill=lightblue, opacity=0.3] fill between[of=upper and lower];		
		\addplot [name path=upper,draw=none] table[x=xi,y expr=\thisrow{JF_max}] {data/simple_w0.2.dat};
		\addplot [name path=lower,draw=none] table[x=xi,y expr=\thisrow{JF_min}] {data/simple_w0.2.dat};
		\addplot [fill=lightred, opacity=0.3] fill between[of=upper and lower];		
		\end{axis}
		\end{tikzpicture}
		\caption{With added process noise $\w = \mathcal{N}(0, 0.04)$}
	\end{subfigure}
	\caption{Comparison of costs achieved for the simple navigation task scenario using Algorithm \ref{alg:pi2solution} with and without funnel adaptation. The robot is modeled with \textbf{single integrator dynamics}. Each value of $\vartheta$ corresponds to a different target cost (\ref{eq:case1Cost}) to be optimized; the plotted $J(\tau)$ costs include a penalty for not satisfying the \ac{STL} task as well. The plotted distributions of the achieved costs were obtained from 20 random runs of the solution algorithm, and show the median of the results. The shaded areas encompass all results excluding the top and bottom $10$th percentiles.}
	\label{fig:resultsIntegrator}
\end{figure*}

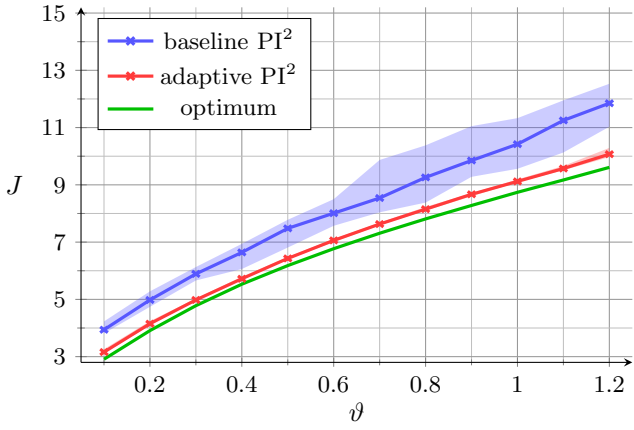
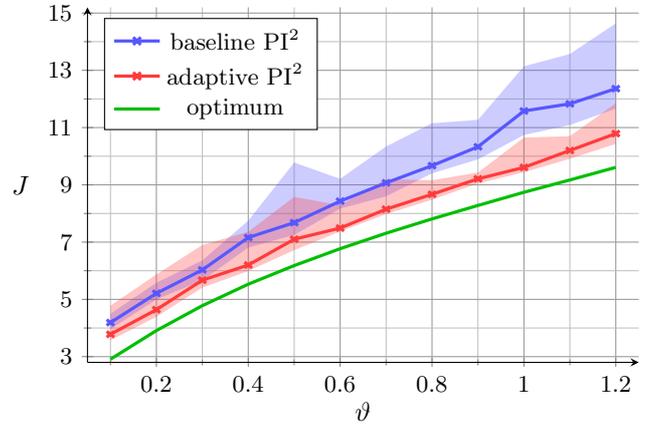
\begin{figure*}[t]
	\centering
	\begin{subfigure}[b]{0.48\linewidth}
		\centering
		\begin{tikzpicture}
		\begin{axis}[grid=both,width=\linewidth,xmin=0.1,ymin=3,ymax=15,xmax=1.2,xlabel=$\vartheta$,ylabel=$J$,legend pos=north west,axis x line=bottom, axis y line = left,ytick={3,5,...,15},ylabel near ticks,ylabel style={rotate=-90},minor x tick num=1, minor y tick num = 1, enlarge x limits={abs=0.05}, enlarge y limits={abs=0.2},grid style={line width=.2pt, draw=gray!50}, major grid style={line width=.4pt,draw=gray!80},yscale=0.78,legend style={font=\small}]
		\addplot [very thick, lightblue, mark=x] table[x=xi,y=JNF] {data/simpleUnicycle_w0.dat};
		\addlegendentry{baseline PI$^2$}
		\addplot [very thick, lightred, mark=x] table[x=xi,y=JF] {data/simpleUnicycle_w0.dat};
		\addlegendentry{adaptive PI$^2$}
		\addplot [very thick, darkgreen, mark=none] table[x=x,y=y] {data/simple_optimum.dat};
		\addlegendentry{optimum}
		
		\addplot [name path=upper,draw=none] table[x=xi,y expr=\thisrow{JNF_max}] {data/simpleUnicycle_w0.dat};
		\addplot [name path=lower,draw=none] table[x=xi,y expr=\thisrow{JNF_min}] {data/simpleUnicycle_w0.dat};
		\addplot [fill=lightblue, opacity=0.3] fill between[of=upper and lower];		
		\addplot [name path=upper,draw=none] table[x=xi,y expr=\thisrow{JF_max}] {data/simpleUnicycle_w0.dat};
		\addplot [name path=lower,draw=none] table[x=xi,y expr=\thisrow{JF_min}] {data/simpleUnicycle_w0.dat};
		\addplot [fill=lightred, opacity=0.3] fill between[of=upper and lower];		
		\end{axis}
		\end{tikzpicture}
		\caption{Without process noise}
	\end{subfigure}
	\begin{subfigure}[b]{0.48\linewidth}
		\centering
		\begin{tikzpicture}
		\begin{axis}[grid=both,width=\linewidth,xmin=0.1,ymin=3,ymax=15,xmax=1.2,xlabel=$\vartheta$,ylabel=$J$,legend pos=north west,axis x line=bottom, axis y line = left,ytick={3,5,...,15},ylabel near ticks,ylabel style={rotate=-90},minor x tick num=1, minor y tick num = 1, enlarge x limits={abs=0.05}, enlarge y limits={abs=0.2},grid style={line width=.2pt, draw=gray!50}, major grid style={line width=.4pt,draw=gray!80},yscale=0.78,legend style={font=\small}]
		\addplot [very thick, lightblue, mark=x] table[x=xi,y=JNF] {data/simpleUnicycle_w0.2.dat};
		\addlegendentry{baseline PI$^2$}
		\addplot [very thick, lightred, mark=x] table[x=xi,y=JF] {data/simpleUnicycle_w0.2.dat};
		\addlegendentry{adaptive PI$^2$}
		\addplot [very thick, darkgreen, mark=none] table[x=x,y=y] {data/simple_optimum.dat};
		\addlegendentry{optimum}
		
		\addplot [name path=upper,draw=none] table[x=xi,y expr=\thisrow{JNF_max}] {data/simpleUnicycle_w0.2.dat};
		\addplot [name path=lower,draw=none] table[x=xi,y expr=\thisrow{JNF_min}] {data/simpleUnicycle_w0.2.dat};
		\addplot [fill=lightblue, opacity=0.3] fill between[of=upper and lower];		
		\addplot [name path=upper,draw=none] table[x=xi,y expr=\thisrow{JF_max}] {data/simpleUnicycle_w0.2.dat};
		\addplot [name path=lower,draw=none] table[x=xi,y expr=\thisrow{JF_min}] {data/simpleUnicycle_w0.2.dat};
		\addplot [fill=lightred, opacity=0.3] fill between[of=upper and lower];		
		\end{axis}
		\end{tikzpicture}
		\caption{With added process noise $\w = \mathcal{N}(0, 0.04)$}
	\end{subfigure}
	\caption{Comparison of costs achieved for the simple navigation task scenario using Algorithm \ref{alg:pi2solution} with and without funnel adaptation. The robot is modeled with \textbf{unicycle dynamics}. Each value of $\vartheta$ corresponds to a different target cost (\ref{eq:case1Cost}) to be optimized; the plotted $J(\tau)$ costs include a penalty for not satisfying the \ac{STL} task as well. The results were obtained in a similar manner as for Figure \ref{fig:resultsIntegrator}. In comparison, due to the nonholonomic dynamics, the guidance controller has more trouble coping with process noise, which is manifests in a higher variance of the achieved results, as seen in (b).}
	\label{fig:resultsUnicycle}
\end{figure*}

Figures \ref{fig:resultsIntegrator} and \ref{fig:resultsUnicycle} present simulation results for solving the outlined scenario using Algorithm \ref{alg:pi2solution} for various values of the target cost parameter $\vartheta$. The two figures correspond to the robot modeled as a single integrator or a unicycle, respectively. For each $\vartheta$, the optimal (penalized) cost $J(\tau)$ can be calculated using (\ref{eq:optCost}) as the penalty term becomes zero when the task is satisfied. The figures show the distribution of the achieved $J(\tau)$ costs obtained from multiple separate runs of the \ac{PI2} algorithm. Results obtained with both funnel adaptation enabled ('adaptive \ac{PI2}') and disabled ('baseline \ac{PI2}') are shown for cases with and without process noise. It is clear that funnel adaptation allows superior results to be achieved; the optimal cost curve is much better approximated and with lower variance.

\begin{figure}[t]
	\centering \hspace{-8mm}
	\begin{tikzpicture}
	\begin{axis}[width=1.0\linewidth,height=0.7\linewidth,grid=both,xmin=-3,ymin=9,ymax=16,xmax=0.25,xlabel=$\rho^{\phi}$,ylabel=$C$,legend pos=north west,axis x line=bottom, axis y line = left,ylabel near ticks,ylabel style={rotate=-90},minor x tick num=1, minor y tick num = 1, enlarge y limits={abs=0.25}, grid style={line width=.2pt, draw=gray!50}, major grid style={line width=.4pt,draw=gray!80},legend style={font=\small}]	
	
	\addplot[only marks,mark=diamond*,green!66!black, opacity=0.8] table[x=rho,y=C] {data/spread_rhoC_adapted.dat};
	\addlegendentry{with adaptation}
	\addplot[only marks,mark=*,orange,opacity=0.8] table[x=rho,y=C] {data/spread_rhoC_vanilla.dat};
	\addlegendentry{without adaptation}
	
	\node (A) at (-1,15.3) {\footnotesize previous iterate};
	\node (B) at (-0.042, 14.5076) {};
	\draw[thick,->] (-0.58,15.1) -- (-0.15, 14.62);
	\draw [decorate,decoration={brace,amplitude=10pt,mirror,raise=4pt},yshift=0pt]
	(-1.1,9.3) -- (0.2,13.4) node [black,midway,xshift=0.75cm,yshift=-0.6cm] {\footnotesize new samples};
	\addplot[mark=*,blue!86!black] (-0.042, 14.5076);

	\end{axis}
	\end{tikzpicture}
	\caption{Distribution of achieved costs and robustness measures during an iteration of \ac{PI2} for solving the simple navigation task with and without funnel adaptation. The results were obtained by allowing the algorithm to run for two iterations without funnel adaptation, then plotting the performance of the trajectories sampled in the next iteration with and without performing a funnel adaptation update beforehand. It is clear that an adapted funnel provides much better guidance towards enforcing the task satisfaction constraint $\rho^{\phi} \ge 0$, allowing more optimal trajectories to be found towards minimizing the target cost $C$ compared to the result of the previous iteration.}
	\label{fig:rhoC}
\end{figure}
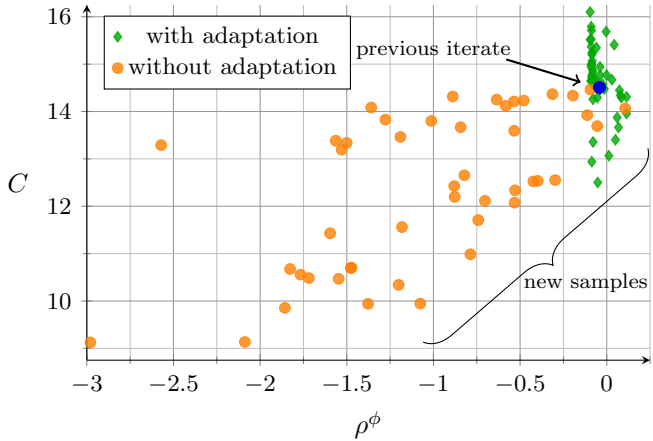

The advantage of funnel adaptation is best illustrated when the initially supplied funnel does not matches the optimal one well, i.e., for higher values of $\vartheta$. In these cases, the robot has to reach and stay at the goal region as early as possible. This is especially difficult to achieve using the open-loop feedforward control terms sought after by the \ac{PI2} algorithm, as exploration or process noise may easily drive the robot away from the goal region after it is reached. Figure \ref{fig:adaptation} demonstrates a sample solution achieved for a target cost defined by $\vartheta = 1.2$, for which the optimal velocity is $v = \min(1,\sqrt{1.2}) = 1.0$ and the optimal time to reach the goal is $T^*_{\text{opt}} = D/v = 4.37$.  A funnel which helps impose such a timing for the evolution of $\rho^{\mu_1}$ (the robustness corresponding to being within the goal area) can greatly aid exploration towards optimal trajectories, and this behavior is exactly what is progressively achieved using funnel adaptation. The figure shows that by the end of the \ac{PI2} iterations, the adapted funnel resembles one that aims to satisfy the \ac{STL} task at hand in a cost-effective manner, i.e., by reaching the goal region near the optimal $T^*_{\text{opt}}$ time. Without a funnel, exploration becomes more difficult as there is no guide controller aiming to keep the robot within the small goal region, and hence the obtained solution is much further from the optimum. This observation is further demonstrated by Fig. \ref{fig:rhoC}.

\begin{figure*}[h!]
	\centering
	\begin{subfigure}[b]{0.48\linewidth}
		\centering
		\begin{tikzpicture}
		\begin{axis}[width=1.0\linewidth,height=0.68\linewidth,grid=both,xmin=0.0,ymin=-5.0,ymax=0.5,xmax=10.4,xlabel=$t$,ylabel=$\rho^{\mu_1}$,legend pos=south east,axis x line=bottom, axis y line = left,ytick={-5,-4,...,0},ylabel near ticks,ylabel style={rotate=-90},minor x tick num=1, minor y tick num = 0, enlarge y limits={abs=0.25}, grid style={line width=.2pt, draw=gray!50}, major grid style={line width=.4pt,draw=gray!80}]	
		
		\addplot [line width=2pt, dashed, lightblue, mark=none] table[x=t,y=rho] {data/simpleSampleSolution_rho0.dat};
		\addlegendentry{$\rho^{\mu_1} \mathrm{(initial)}$}
		\addplot [line width=2pt, dashed, lightred, mark=none] table[x=t,y=gamma] {data/simpleSampleSolution_rho0.dat};
		\addlegendentry{$\gamma_1 \mathrm{(initial)}$}
		\addplot [line width=2pt, lightblue, mark=none] table[x=t,y=rho] {data/simpleSampleSolution_rho.dat};
		\addlegendentry{$\rho^{\mu_1} \mathrm{(final)}$}
		\addplot [line width=2pt, lightred, mark=none] table[x=t,y=gamma] {data/simpleSampleSolution_rho.dat};
		\addlegendentry{$\gamma_1 \mathrm{(final)}$}
		
		\draw [thick,color=black] (4.368,-6) -- (4.368,0.5);
		\draw [thick,color=black] (4.9,-6) -- (4.9,0.5);
		\node [fill=white, rounded corners=2pt] at (3.0,0.5) {\small \color{black}$T^*_{\text{opt}} = 4.37$};
		\node [fill=white, rounded corners=2pt] at (6.25,0.55) {\small \color{black}$T^* = 4.90$};
		
		\end{axis}
		\end{tikzpicture}
		\caption{With funnel adaptation, $J(\tau) = 10.18$}
	\end{subfigure} \hfill
	\begin{subfigure}[b]{0.48\linewidth}
		\centering
		\begin{tikzpicture}
		\begin{axis}[width=1.0\linewidth,height=0.68\linewidth,grid=both,xmin=0.0,ymin=-5.0,ymax=0.5,xmax=10.4,xlabel=$t$,ylabel=$\rho^{\mu_1}$,legend pos=south east,axis x line=bottom, axis y line = left,ytick={-5,-4,...,0},ylabel near ticks,ylabel style={rotate=-90},minor x tick num=1, minor y tick num = 0, enlarge y limits={abs=0.25}, grid style={line width=.2pt, draw=gray!50}, major grid style={line width=.4pt,draw=gray!80}]	
		
		\addplot [line width=2pt, dashed, lightblue, mark=none] table[x=t,y=rho] {data/simpleSampleSolution_rho0.dat};
		\addlegendentry{$\rho^{\mu_1} \mathrm{(initial)}$}
		\addplot [line width=2pt, lightblue, mark=none] table[x=t,y=rho] {data/simpleSampleSolution_rho_vanilla.dat};
		\addlegendentry{$\rho^{\mu_1} \mathrm{(final)}$}
		\addplot [line width=2pt, lightred, mark=none] table[x=t,y=gamma] {data/simpleSampleSolution_rho_vanilla.dat};
		\addlegendentry{$\gamma_1$}
		
		\draw [thick,color=black] (4.368,-6) -- (4.368,0.5);
		\draw [thick,color=black] (9.82,-6) -- (9.82,0.5);
		\node [fill=white, rounded corners=2pt] at (3.0,0.5) {\small \color{black}$T^*_{\text{opt}} = 4.37$};
		\node [fill=white, rounded corners=2pt] at (8.55,0.55) {\small \color{black}$T^* = 9.82$};
		
		\end{axis}
		\end{tikzpicture}
		\caption{Without funnel adaptation, $J(\tau) = 14.6$}
	\end{subfigure}
	\caption{Illustration of funnel adaptation for the simple navigation scenario for a target cost defined by $\vartheta = 1.2$. Optimally, the robustness $\rho^{\mu_1}$ corresponding to being within the goal region should be reached at $T^*_{\text{opt}} = 4.37$s. The adapted funnel successfully enforces $T^*=4.90$s, which is much closer to the optimum than $T^* = 9.82$s achieved without funnel adaptation.}
	\label{fig:adaptation}
\end{figure*}
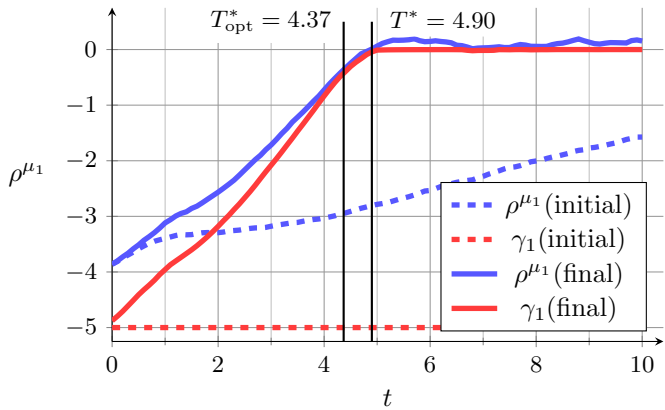
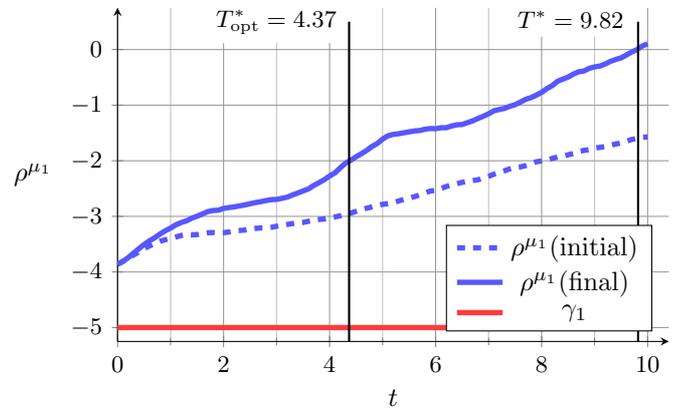

A final observation is that similar quality solutions were found for both the single integrator and unicycle model systems. This is especially notable considering that this was accomplished by keeping all of the \ac{PI2} algorithm and funnel adaptation hyperparameters the same for the two cases; only the guiding controllers were changed according to the different system dynamics. Without the introduced improved combination controller and funnel adaptation scheme, more problem-dependent tuning was required to achieve optimal results, especially for more complex scenarios such as the one examined next.

\subsection{Complex navigation task}

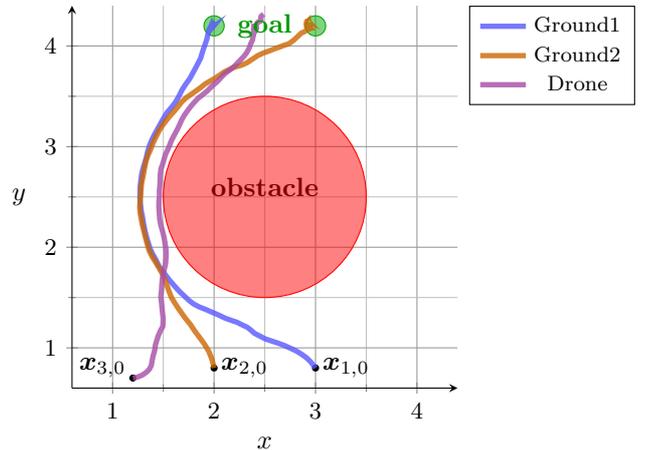
\begin{figure}[t]
	\centering
	\begin{tikzpicture}
	\begin{axis}[width=0.74\linewidth, height=0.74\linewidth,grid=both,legend style={font=\footnotesize},legend pos=outer north east, axis equal,xmin=1,ymin=1,ymax=4,xmax=4,xlabel=$x$,ylabel=$y$,axis x line=bottom, axis y line = left,ytick={1,...,4},xtick={1,...,4},ylabel near ticks,ylabel style={rotate=-90},minor x tick num=1, minor y tick num = 1,grid style={line width=.2pt, draw=gray!50}, enlarge y limits={abs=0.4}, enlarge x limits={abs=0.4}, major grid style={line width=.4pt,draw=gray!80}]
	
	\filldraw[fill=black,draw=black] (1.2,0.7) circle (.03);
	\filldraw[fill=black,draw=black] (2.0,0.8) circle (.03);
	\filldraw[fill=black,draw=black] (3.0,0.8) circle (.03);
	\filldraw[fill=red,draw=red,fill opacity=0.5] (2.5,2.5) circle (1);
	\filldraw[fill=green!66!black,draw=green!66!black,fill opacity=0.5] (2.0,4.2) circle (0.1);
	\filldraw[fill=green!66!black,draw=green!66!black,fill opacity=0.5] (3.0,4.2) circle (0.1);
	
	\addplot [line width=2pt, lightblue, mark=none, opacity=0.8] table[x=x1,y=x2] {data/complex/sampleSolution.dat};
	\addlegendentry{Ground1}
	\addplot [line width=2pt, orange!80!black, mark=none, opacity=0.8] table[x=x3,y=x4] {data/complex/sampleSolution.dat};
	\addlegendentry{Ground2}
	\addplot [line width=2pt, red!50!blue!70!white, mark=none, opacity=0.8] table[x=x5,y=x6] {data/complex/sampleSolution.dat};
	\addlegendentry{Drone}
	
	\node at (2.5,2.6) {\color{red!50!black} \textbf{obstacle}};
	\node at (2.5,4.2) {\color{green!60!black} \textbf{goal}};
	\node at (0.9,0.8) {$\x[3,0]$};
	\node at (2.3,0.8) {$\x[2,0]$};
	\node at (3.3,0.8) {$\x[1,0]$};
	\end{axis}
	\end{tikzpicture}
	\caption{Depiction of the complex navigation task scenario, along with a sample trajectory for solving the \ac{STL} task. The ground robots must eventually reach and stay at respective goal regions while maintaining a roughly fixed distance between themselves. The drone has to eventually reach and stay at the center of the two robots.}
	\label{fig:case2Sample}
\end{figure}

This section presents a more elaborate scenario illustrating the applicability of the developed policy search algorithm for solving \ac{STL} tasks. We show how funnel adaptation allows near-optimal solutions to be found even for a more complicated task and higher dimensional system than in the simple navigation task example. The scenario involves two ground robots which must eventually reach and stay at target locations while avoiding an obstacle and maintaining the distance between themselves within given bounds. Furthermore, a drone has to eventually reach and follow the center of the two robots. The robots must accomplish this task while minimizing their input energy during the problem time horizon of $T=10$s. The scenario and a sample solution is shown in Figure \ref{fig:case2Sample}. The detailed scenario description, along with the \ac{PI2} algorithm parameters used for finding a solution are given in Appendix \ref{appendix:complexScenario}. 

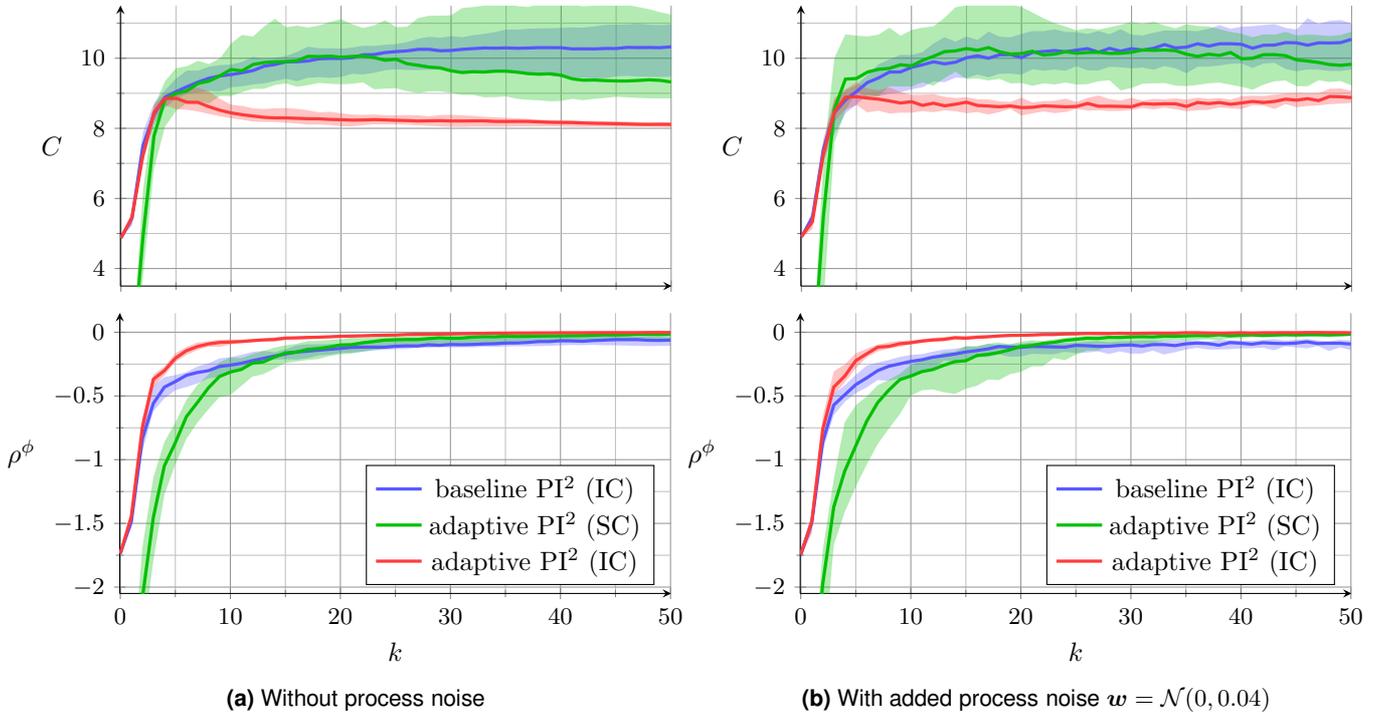
\begin{figure*}[h!]
	\centering
	\begin{subfigure}[b]{0.48\linewidth}
		\centering
		\begin{tikzpicture}
		\begin{axis}[grid=both,width=\linewidth,height=0.6\linewidth,xmin=0,ymin=4,ymax=11,xmax=50,ylabel=$C$,legend pos=south east,axis x line=bottom, axis y line = left,ylabel near ticks,ylabel style={rotate=-90},xticklabels={,,},minor x tick num=1, minor y tick num = 1, enlarge x limits={abs=0.05}, enlarge y limits={abs=0.5},grid style={line width=.2pt, draw=gray!50}, major grid style={line width=.4pt,draw=gray!80}]
		\addplot [very thick, lightblue, mark=none] table[x=k,y=C] {data/complex/w0_guideless_vanilla.dat};
		\addplot [very thick, darkgreen, mark=none] table[x=k,y=C] {data/complex/w0_guideless_simple0.8.dat};
		\addplot [very thick, lightred, mark=none] table[x=k,y=C] {data/complex/w0_guideless_improved0.8.dat};
		
		\addplot [name path=upper,draw=none] table[x=k,y expr=\thisrow{Cmax}] {data/complex/w0_guideless_vanilla.dat};
		\addplot [name path=lower,draw=none] table[x=k,y expr=\thisrow{Cmin}] {data/complex/w0_guideless_vanilla.dat};
		\addplot [fill=lightblue, opacity=0.3] fill between[of=upper and lower];		
		\addplot [name path=upper,draw=none] table[x=k,y expr=\thisrow{Cmax}] {data/complex/w0_guideless_simple0.8.dat};
		\addplot [name path=lower,draw=none] table[x=k,y expr=\thisrow{Cmin}] {data/complex/w0_guideless_simple0.8.dat};
		\addplot [fill=darkgreen, opacity=0.3] fill between[of=upper and lower];
		\addplot [name path=upper,draw=none] table[x=k,y expr=\thisrow{Cmax}] {data/complex/w0_guideless_improved0.8.dat};
		\addplot [name path=lower,draw=none] table[x=k,y expr=\thisrow{Cmin}] {data/complex/w0_guideless_improved0.8.dat};
		\addplot [fill=lightred, opacity=0.3] fill between[of=upper and lower];		
		\end{axis}
		\end{tikzpicture} \\ \hspace{-3.7mm}
		\begin{tikzpicture}
		\begin{axis}[grid=both,width=\linewidth,height=0.6\linewidth,xmin=0,ymin=-2,ymax=0.1,xmax=50,xlabel=$k$,ylabel=$\rho^{\phi}$,legend pos=south east,axis x line=bottom, axis y line = left,ylabel near ticks,ylabel style={rotate=-90},minor x tick num=1, minor y tick num = 1, enlarge x limits={abs=0.05}, enlarge y limits={abs=0.05},grid style={line width=.2pt, draw=gray!50}, major grid style={line width=.4pt,draw=gray!80}]
		\addplot [very thick, lightblue, mark=none] table[x=k,y=rho] {data/complex/w0_guideless_vanilla.dat};
		\addlegendentry{baseline PI$^2$ (IC)}
		\addplot [very thick, darkgreen, mark=none] table[x=k,y=rho] {data/complex/w0_guideless_simple0.8.dat};
		\addlegendentry{adaptive PI$^2$ (SC)}
		\addplot [very thick, lightred, mark=none] table[x=k,y=rho] {data/complex/w0_guideless_improved0.8.dat};
		\addlegendentry{adaptive PI$^2$ (IC)}
		
		\addplot [name path=upper,draw=none] table[x=k,y expr=\thisrow{rhomax}] {data/complex/w0_guideless_vanilla.dat};
		\addplot [name path=lower,draw=none] table[x=k,y expr=\thisrow{rhomin}] {data/complex/w0_guideless_vanilla.dat};
		\addplot [fill=lightblue, opacity=0.3] fill between[of=upper and lower];		
		\addplot [name path=upper,draw=none] table[x=k,y expr=\thisrow{rhomax}] {data/complex/w0_guideless_simple0.8.dat};
		\addplot [name path=lower,draw=none] table[x=k,y expr=\thisrow{rhomin}] {data/complex/w0_guideless_simple0.8.dat};
		\addplot [fill=darkgreen, opacity=0.3] fill between[of=upper and lower];
		\addplot [name path=upper,draw=none] table[x=k,y expr=\thisrow{rhomax}] {data/complex/w0_guideless_improved0.8.dat};
		\addplot [name path=lower,draw=none] table[x=k,y expr=\thisrow{rhomin}] {data/complex/w0_guideless_improved0.8.dat};
		\addplot [fill=lightred, opacity=0.3] fill between[of=upper and lower];		
		
		\end{axis}
		\end{tikzpicture}
		\caption{Without process noise}
	\end{subfigure}
	\begin{subfigure}[b]{0.48\linewidth}
		\centering
		\begin{tikzpicture}
		\begin{axis}[grid=both,width=\linewidth,height=0.6\linewidth,xmin=0,ymin=4,ymax=11,xmax=50,ylabel=$C$,legend pos=south east,axis x line=bottom, axis y line = left,ylabel near ticks,ylabel style={rotate=-90},xticklabels={,,},minor x tick num=1, minor y tick num = 1, enlarge x limits={abs=0.05}, enlarge y limits={abs=0.5},grid style={line width=.2pt, draw=gray!50}, major grid style={line width=.4pt,draw=gray!80}]
		\addplot [very thick, lightblue, mark=none] table[x=k,y=C] {data/complex/w0.2_guideless_vanilla.dat};
		\addplot [very thick, darkgreen, mark=none] table[x=k,y=C] {data/complex/w0.2_guideless_simple0.8.dat};
		\addplot [very thick, lightred, mark=none] table[x=k,y=C] {data/complex/w0.2_guideless_improved0.8.dat};
		
		\addplot [name path=upper,draw=none] table[x=k,y expr=\thisrow{Cmax}] {data/complex/w0.2_guideless_vanilla.dat};
		\addplot [name path=lower,draw=none] table[x=k,y expr=\thisrow{Cmin}] {data/complex/w0.2_guideless_vanilla.dat};
		\addplot [fill=lightblue, opacity=0.3] fill between[of=upper and lower];		
		\addplot [name path=upper,draw=none] table[x=k,y expr=\thisrow{Cmax}] {data/complex/w0.2_guideless_simple0.8.dat};
		\addplot [name path=lower,draw=none] table[x=k,y expr=\thisrow{Cmin}] {data/complex/w0.2_guideless_simple0.8.dat};
		\addplot [fill=darkgreen, opacity=0.3] fill between[of=upper and lower];
		\addplot [name path=upper,draw=none] table[x=k,y expr=\thisrow{Cmax}] {data/complex/w0.2_guideless_improved0.8.dat};
		\addplot [name path=lower,draw=none] table[x=k,y expr=\thisrow{Cmin}] {data/complex/w0.2_guideless_improved0.8.dat};
		\addplot [fill=lightred, opacity=0.3] fill between[of=upper and lower];		
		\end{axis}
		\end{tikzpicture} \\ \hspace{-3.7mm}
		\begin{tikzpicture}
		\begin{axis}[grid=both,width=\linewidth,height=0.6\linewidth,xmin=0,ymin=-2,ymax=0.1,xmax=50,xlabel=$k$,ylabel=$\rho^{\phi}$,legend pos=south east,axis x line=bottom, axis y line = left,ylabel near ticks,ylabel style={rotate=-90},minor x tick num=1, minor y tick num = 1, enlarge x limits={abs=0.05}, enlarge y limits={abs=0.05},grid style={line width=.2pt, draw=gray!50}, major grid style={line width=.4pt,draw=gray!80}]
		\addplot [very thick, lightblue, mark=none] table[x=k,y=rho] {data/complex/w0.2_guideless_vanilla.dat};
		\addlegendentry{baseline PI$^2$ (IC)}
		\addplot [very thick, darkgreen, mark=none] table[x=k,y=rho] {data/complex/w0.2_guideless_simple0.8.dat};
		\addlegendentry{adaptive PI$^2$ (SC)}
		\addplot [very thick, lightred, mark=none] table[x=k,y=rho] {data/complex/w0.2_guideless_improved0.8.dat};
		\addlegendentry{adaptive PI$^2$ (IC)}
		
		\addplot [name path=upper,draw=none] table[x=k,y expr=\thisrow{rhomax}] {data/complex/w0.2_guideless_vanilla.dat};
		\addplot [name path=lower,draw=none] table[x=k,y expr=\thisrow{rhomin}] {data/complex/w0.2_guideless_vanilla.dat};
		\addplot [fill=lightblue, opacity=0.3] fill between[of=upper and lower];		
		\addplot [name path=upper,draw=none] table[x=k,y expr=\thisrow{rhomax}] {data/complex/w0.2_guideless_simple0.8.dat};
		\addplot [name path=lower,draw=none] table[x=k,y expr=\thisrow{rhomin}] {data/complex/w0.2_guideless_simple0.8.dat};
		\addplot [fill=darkgreen, opacity=0.3] fill between[of=upper and lower];
		\addplot [name path=upper,draw=none] table[x=k,y expr=\thisrow{rhomax}] {data/complex/w0.2_guideless_improved0.8.dat};
		\addplot [name path=lower,draw=none] table[x=k,y expr=\thisrow{rhomin}] {data/complex/w0.2_guideless_improved0.8.dat};
		\addplot [fill=lightred, opacity=0.3] fill between[of=upper and lower];		
		\end{axis}
		\end{tikzpicture}
		\caption{With added process noise $\w = \mathcal{N}(0, 0.04)$}
	\end{subfigure}
	\caption{Convergence of the target cost $C$ and task satisfaction robustness metric $\rho^{\phi}$ for solving the complex scenario. The graphs show the median achieved values during $20$ sample runs of the \ac{PI2} algorithm as a function of the iteration number $k$. Results are presented with and without funnel adaptation active; for the former case, using the simple combination (`SC') controller is also shown in order to illustrate the effectiveness of the improved combination (`IC') controller.}
	\label{fig:complex}
\end{figure*}

Figure \ref{fig:complex} shows the convergence rate of \ac{PI2} for different variations of Algorithm \ref{alg:pi2solution}, with and without process noise affecting the system. The `baseline' and `adaptive' tags denote whether funnel adaptation is turned off or on. The `SC' and `IC' tags refer to whether the simple (\ref{eq:simpleCombination}) or the improved (\ref{eq:improvedCombination}) combination controllers were used as guidance laws. The results illustrate that the algorithm has difficulty satisfying the \ac{STL} task specification without funnel adaptation, especially in the presence of process noise. On the other hand, task satisfaction is achieved with funnel adaptation, and the improved combination controller furthermore allows more optimal trajectories to be found with respect to the target cost $C$. The benefits of using this latter controller are also seen in terms of the increased convergence rate, and thus increased sample efficiency of the algorithm.

We note that the cost $C(\tau) \approx 8$ was also achieved in \cite{varnai2019prescribed}, without funnel adaptation or an improved combination controller. However, this required much more elaborate and problem-specific tuning of the guidance controller, whereas the current results were essentially obtained using the same algorithm hyperparameters as in the simple scenario example.

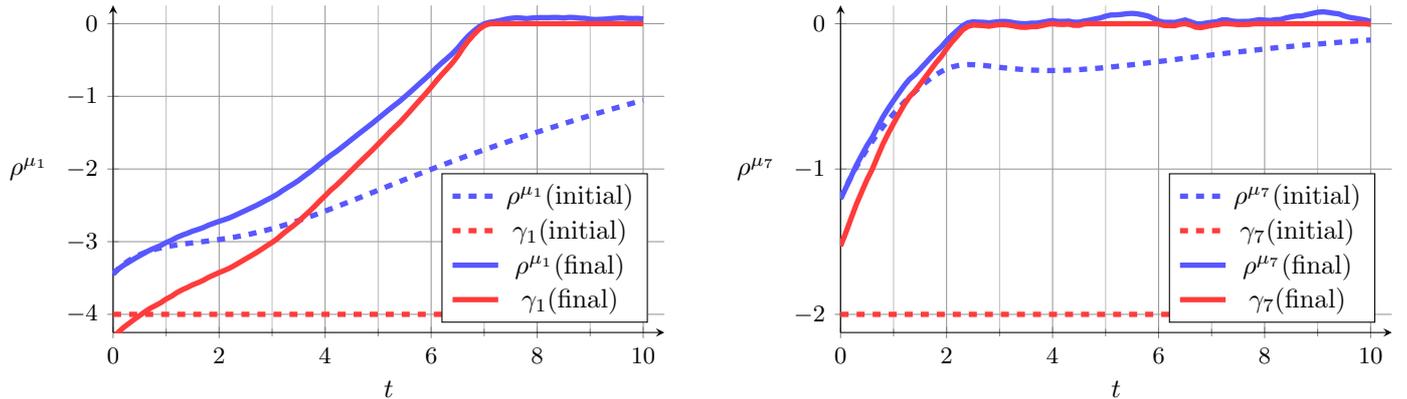
\begin{figure*}[h!]
	\centering
	\begin{subfigure}{0.48\linewidth}
		\centering
		\begin{tikzpicture}
		\begin{axis}[width=1.0\linewidth,height=0.67\linewidth,grid=both,xmin=0.0,ymin=-4.0,ymax=0.0,xmax=10.4,xlabel=$t$,ylabel=$\rho^{\mu_1}$,legend pos=south east,axis x line=bottom, axis y line = left,ytick={-4,-3,...,0},ylabel near ticks,ylabel style={rotate=-90},minor x tick num=1, minor y tick num = 0, enlarge y limits={abs=0.25}, grid style={line width=.2pt, draw=gray!50}, major grid style={line width=.4pt,draw=gray!80}]	
		
		\addplot [line width=2pt, dashed, lightblue, mark=none] table[x=t,y=rho10] {data/complex/sampleMeasures.dat};
		\addlegendentry{$\rho^{\mu_1} \mathrm{(initial)}$}
		\addplot [line width=2pt, dashed, lightred, mark=none] table[x=t,y=gamma10] {data/complex/sampleMeasures.dat};
		\addlegendentry{$\gamma_1 \mathrm{(initial)}$}
		\addplot [line width=2pt, lightblue, mark=none] table[x=t,y=rho1] {data/complex/sampleMeasures.dat};
		\addlegendentry{$\rho^{\mu_1} \mathrm{(final)}$}
		\addplot [line width=2pt, lightred, mark=none] table[x=t,y=gamma1] {data/complex/sampleMeasures.dat};
		\addlegendentry{$\gamma_1 \mathrm{(final)}$}
		
		\end{axis}
		\end{tikzpicture}
	\end{subfigure} \hfill
	\begin{subfigure}{0.48\linewidth}
		\centering
		\begin{tikzpicture}
		\begin{axis}[width=1.0\linewidth,height=0.67\linewidth,grid=both,xmin=0.0,ymin=-2,ymax=0.0,xmax=10.4,xlabel=$t$,ylabel=$\rho^{\mu_7}$,legend pos=south east,axis x line=bottom, axis y line = left,ytick={-5,-4,...,0},ylabel near ticks,ylabel style={rotate=-90},minor x tick num=1, minor y tick num = 0, enlarge y limits={abs=0.125}, grid style={line width=.2pt, draw=gray!50}, major grid style={line width=.4pt,draw=gray!80}]	
		
		\addplot [line width=2pt, dashed, lightblue, mark=none] table[x=t,y=rho70] {data/complex/sampleMeasures.dat};
		\addlegendentry{$\rho^{\mu_7} \mathrm{(initial)}$}
		\addplot [line width=2pt, dashed, lightred, mark=none] table[x=t,y=gamma70] {data/complex/sampleMeasures.dat};
		\addlegendentry{$\gamma_7 \mathrm{(initial)}$}
		\addplot [line width=2pt, lightblue, mark=none] table[x=t,y=rho7] {data/complex/sampleMeasures.dat};
		\addlegendentry{$\rho^{\mu_7} \mathrm{(final)}$}
		\addplot [line width=2pt, lightred, mark=none] table[x=t,y=gamma7] {data/complex/sampleMeasures.dat};
		\addlegendentry{$\gamma_7 \mathrm{(final)}$}
		
		\end{axis}
		\end{tikzpicture}
	\end{subfigure}
		\caption{Funnel adaptation in action for the complex scenario for a sample case where the user does not have an estimate of funnels which enforce task satisfaction. The task requires $\mu_1$ and $\mu_7$ to become true at latest $7$s and $3$s, respectively. As the input energy should be minimized, these are indeed the times around which the actual metrics obtain positive robustness.\vspace{-2mm}}
		\label{fig:complexFunnelAdaptation}
\end{figure*}

Figure \ref{fig:complexFunnelAdaptation} shows how the funnels corresponding to reaching the target region with the first ground robot (predicate $\mu_1$) and to reaching the middle of the ground robots with the drone (predicate $\mu_7$) are adapted during a sample solution process. As with the simple scenario, here we also see that the final funnels follow the optimal robustness evolutions dictated by the task definition and target cost to minimize. For example, the task requires the goal region to be reached within $7$s, and in order to minimize the expended input energy, this is the latest possible time at which it should be reached. The funnel corresponding to $\mu_1$ indeed aims to enforce such a behavior.

	\section{Conclusions} \label{section:conclusions}

In this work, we examined the applicability of the \ac{PI2} learning method for controlling systems under \ac{STL} task specifications while minimizing a target cost of interest. We introduced a controller derivation framework inspired by general learning methods, termed the penalty-based framework, for deriving controllers that, while do not guarantee task satisfaction, give good guidance towards it in order to effectively aid exploration during the learning procedure. We also proposed a funnel adaptation scheme which updates the parameters of the guidance controller in order to maintain its relevance and improve its guidance as the iterations progress.

While the results are quite promising, the detailed algorithm leaves much room for further improvement. For example, recent years have seen many definitions of quantitative robustness metrics for \ac{STL} formulas. Some of these measures might be better suited as a reward for the \ac{PI2} learning procedure. Reward shaping is known to be an important aspect for the performance of learning methods, and while there is a drawback of requiring domain knowledge, it may still be possible to engineer a robustness definition that performs well holistically across a wide array of practically relevant \ac{STL} task specifications, and is thus of considerable interest for future work. Another crucial aspect of the proposed algorithm is its role as a general sampling-based method for solving a constrained optimization problems. So far, the task satisfaction constraint has been incorporated into the objective to minimize through a penalty function and was enforced by progressively increasing the amount of penalty imposed on task violation. This increasement has a significant impact on the algorithm's performance, and it would be greatly beneficial to provide theoretically founded rules for determining it instead of treating it as a hyperparameter to be tuned.
	
	\vspace{6mm}
	\noindent
	{\bfseries Acknowledgements}
	
	\vspace{1mm}
	\noindent
	This work was partially supported by the Wallenberg AI, Autonomous Systems and Software Program (WASP) funded by the Knut and Alice Wallenberg Foundation, the Swedish Research Council (VR), the SSF COIN project, and the EU H2020 Co4Robots project.

	\bibliographystyle{plainnat}
	\bibliography{MyBib}
	
	\newpage
	

	\appendix

\section{Theorem proofs}

The main proofs of the developed theoretical framework make use of the following lemmas.
 
\begin{lemma}[Theorem 3.1, Local Existence \& Uniqueness \citep{khalil2002nonlinear}] \label{lemma:odeSolution}
	Consider the initial value problem $\x<\dot> = f(\x, t)$ with given $\x(t_0) = \x[0]$. Suppose the function $f$ is uniformly Lipschitz continuous in $\x$ and piecewise continuous in $t$ in a closed ball $\mathcal{B} = \left\{\x \inReal[n], t \inReal : \norm{\x - \x[0]} \le r,\ t \in [t_0,\ t_1]\right\}$. Then, there exists some $\delta > 0$ such that the initial value problem has a unique solution over the time interval $[t_0, t_0 + \delta]$.
\end{lemma}
 
\begin{lemma}[Generalized Nagumo's Theorem, {\cite[Section 4.2.2]{blanchini2015set}}] \label{lemma:nagumo}
 	Consider the system $\x<\dot> = f(\x, t)$ and time-varying sets of the form $S(t) = \left\{\x : \zeta(\x, t) \le 0 \right\}$ where $\zeta(\x, t)$ is smooth. Assume that the system admits a unique solution and that at any $t$ we have $\frac{\partial \zeta(\x, t)}{\partial \x} \ne \nvec$ for $\zeta(\x, t) = 0$. Then, the condition $x(t') \in S(t')$ implies $x(t) \in S(t)$ for $t \ge t'$ if the inequality $\dot{\zeta}(\x, t) \le 0$ holds at the boundary $\zeta(\x, t) = 0$.
\end{lemma}

\subsection{Proof of Theorem \ref{theorem:localSatisfaction}}

The proof is based on the one given in \cite{varnai2019prescribed}.

\begin{proof}
	Let the system at time $t'$ be at a state $\x(t')$ for which $\rho^{\mu}(\x(t')) \ge \gamma(t')$ holds. To prove local robustness satisfaction, we show that under the defined control law (\ref{eq:baseControlFamily}), a unique solution exists and $\rho^{\mu}(\x(t)) \ge \gamma(t)$ remains satisfied during $t \in [t', t' + \delta]$ for some $\delta > 0$ period of time. For the former, in order to apply Lemma \ref{lemma:odeSolution}, we must show that there exists a closed ball around $(\x(t'), t')$ within which $\u(\x, t)$ is Lipschitz continuous in $\x$ and piecewise continuous in $t$. Then the same holds for $\f(\x) + \g(\x)\u(\x, t) + \w(t)$, the right hand side of (\ref{eq:systemDynamics}), due to Assumption \ref{assumptions:dynamics}, and the lemma can be applied. 
	
	Piecewise continuity in $t$ trivially holds due to the continuity of $\kappa(\x, t)$ and $\mathcal{A}(t)$ in $t$. The Lipschitz condition also holds trivially for any $\x(t') \in \mathcal{A}(t)$ where the control is defined to be zero. If $\x(t') \notin \mathcal{A}(t')$, then we must have $\x(t') \in \mathcal{X}(t')$ for which $\norm[2]{\v(\x(t'))} > 0$ by Assumption \ref{assumption:controllability}. Thus, as $\v(\x)$ is continuous, there exists a closed ball $\mathcal{B}$ around $(\x(t'), t')$ in which $\norm[2]{\v(\x)}$ is nonzero. Furthermore, as $\v(\x)$ and $\kappa(\x,t)$ are locally Lipschitz, the control action (\ref{eq:baseControlFamily}) is also locally Lipschitz in $\mathcal{B}$. The Lipschitz property of $\u(\x, t)$ is preserved at the boundary $\bar{\mathcal{X}}(t)$ where $\u$ is continuous. Therefore, Lemma \ref{lemma:odeSolution} is applicable and a unique solution exists for some time interval $t \in [t', t' + \delta]$ with $\delta > 0$ from the initial condition $\x(t')$.
	
	The proof of local robustness satisfaction is completed by showing that during this time $\rho^{\mu}(\x(t)) \ge \gamma(t)$ remains true (for any time interval, in fact, for which a solution exists; i.e., the set is forward invariant). A sufficient condition for this is given by extensions of Nagumo's Theorem (see Lemma \ref{lemma:nagumo}). Applying the lemma to the set defined as $S(t) = \left\{\x: \gamma(t) - \rho^{\mu}(\x) \le 0 \right\}$ yields the condition:
	\begin{equation} \label{eq:satisfactionCondition}
	\dot{\rho}^{\mu}(\x) \ge \dot{\gamma}(t) \quad \text{if } \x \in \lbar{\mathcal{X}}(t),
	\end{equation} 
	which, if satisfied, implies that the trajectory of $\rho^{\mu}(\x)$, having started above $\gamma(t)$, cannot cross it, as desired. Let the controller parameters be $K = 1$ and $\varDelta = 0$, or, if a lower bound $\norm[2]{\v(\x)} \ge v_{\text{min}} > 0$ is known for all $\x \in \mathcal{X}(t)$ uniformly across $t$, satisfy $(K - 1)v^2_{\text{min}} \ge \varDelta$. With either choice, we also have $(K - 1)\norm[2]{\v(\x)}^2 \ge \varDelta$ for all $\x \in \mathcal{X}(t)$, thus the inequality
	\begin{equation}
	\dfrac{K}{\norm[2]{\v(\x)}^2 + \varDelta} \ge \dfrac{1}{\norm[2]{\v(\x)}^2}
	\end{equation}
	holds in this set as well. Substituting the control law (\ref{eq:baseControlFamily}) at $\x \in \lbar{\mathcal{X}}(t)$ into the time derivative (\ref{eq:robustnessDerivative}) of $\rho^{\mu}(\x(t))$, and using the condition (i) imposed on $\kappa(\x, t)$ by Theorem \ref{theorem:localSatisfaction}, we can show that Nagumo's condition is then satisfied at the required $\x \in \lbar{\mathcal{X}}(t)$ region:
	\begingroup
	\allowdisplaybreaks
	\begin{align*}
	\dot{\rho}^{\mu}(\x) &= \dfrac{\partial \rho^{\mu}(\x)}{\partial \x} \left(\f(\x) + \w\right) + \v(\x)\tp \dfrac{\kappa(\x, t)K}{\norm[2]{\v(\x)}^{2} + \varDelta}\v(\x) \\
	&\ge \dfrac{\partial \rho^{\mu}(\x)}{\partial \x} \left(\f(\x) + \w\right) + \dfrac{\kappa(\x, t)}{\norm[2]{\v(\x)}^{2}}\v(\x)\tp \v(\x) \\
	&\ge \dfrac{\partial \rho^{\mu}(\x)}{\partial \x} \f(\x) - \max_{\w}\norm{\dfrac{\partial \rho^{\mu}(\x)}{\partial \x} \w} \\
	&\phantom{\ge}+ \dot{\gamma}(t) - \dfrac{\partial \rho^{\mu}(\x)}{\partial \x} \f(\x) + \max_{\w}\norm{\dfrac{\partial \rho^{\mu}(\x)}{\partial \x} \w} \\
	&= \dot{\gamma}(t),
	\end{align*}
	\endgroup
	as was to be shown for local robustness satisfaction.
\end{proof}

\subsection{Proof of Theorem \ref{theorem:penaltyController}}
\begin{proof}
	We begin by expanding the derivative $\dot{P}$ in the cost $J(\u)$ from the optimization problem (\ref{eq:singlePenaltyOptimization}) as:
	\begin{align*}
		J(\u) = &\dfrac{\partial P(\rho^{\mu}(\x), t)}{\partial \rho^{\mu}}\dfrac{\partial \rho^{\mu}(\x)}{\partial \x}\x<\dot> + \dfrac{\partial P(\rho^{\mu}(\x), t)}{\partial t} \\ &+ \dfrac{1}{2}\u\tp \R(\x) \u.
	\end{align*}
	Substituting in the system dynamics (\ref{eq:systemDynamics_restated2}) for $\x<\dot>$ and keeping the terms involving the input, we can see that minimizing $J(\u)$ is equivalent to minimizing the term:
	\begin{equation} \label{eq:equivalentOpt}
		\dfrac{\partial P(\rho^{\mu}(\x), t)}{\partial \rho^{\mu}}\v(\x)\tp \u + \dfrac{1}{2}\u\tp \R(\x) \u,
	\end{equation}
	where $\v(\x)\tp = \frac{\partial \rho^{\mu}(\x)}{\partial \x} g(\x)$, as defined previously. The minimizer to this expression is obtained by setting the gradient with respect to the input to zero, yielding:
	\begin{equation} \label{eq:penaltyController}
		\u(\x, t) = -\dfrac{\partial P(\rho^{\mu}(\x), t)}{\partial \rho^{\mu}}\R(\x)\inv\v(\x).
	\end{equation}
	The proof essentially follows by substituting in the choices of $\R(\x)$ into the derived solution (\ref{eq:penaltyController}), and showing that the controllers have the same form as the previously derived (\ref{eq:baseControlFamily}). Briefly, $\R(\x) = \R{'}(\x) = K^{-1}(\norm[2]{\v(\x)}^2 + \Delta)\I$ immediately gives the desired:
	\begin{equation}
	\u(\x, t) = -\dfrac{\partial P(\rho^{\mu}(\x), t)}{\partial \rho^{\mu}}\dfrac{K}{\norm[2]{\v(\x)}^2 + \Delta}\v(\x).
	\end{equation}
	On the other hand, with $\R(\x) = \R{''}(\x) = K^{-1}\left(\v(\x)\v(\x)\tp + \Delta \I\right)$, the inverse of this term can be computed using the Sherman-Morrison formula to yield:
	\begin{align*}
	\R{''}\inv &= K(\Delta \I + \v\v\tp)\inv \\
	&= K\Delta\inv \I - K\Delta\inv \dfrac{\v \v\tp}{1 + \v\tp \Delta\inv \v } \Delta\inv \\
	&= K\Delta\inv \left(\I - \dfrac{\v \v\tp}{\Delta + \v\tp \v }\right).
	\end{align*}
	In the input (\ref{eq:penaltyController}), when multiplied from the right by $\v$, this becomes:
	\begin{align*}
	\R{''}\inv\v &= K\Delta\inv \left(\I - \dfrac{\v \v\tp}{\Delta + \v\tp \v }\right) \v \\
	&= K\Delta\inv \left(\v - \dfrac{\v \v\tp \v}{\Delta + \v\tp \v }\right) \\
	&= K\Delta\inv \left(\dfrac{\v(\Delta + \v\tp \v) - \v \v\tp \v}{\Delta + \v\tp \v }\right) \\
	&= K\Delta\inv \left(\dfrac{\Delta\v}{\Delta + \v\tp \v }\right) \\
	&= \dfrac{K}{\norm[2]{\v}^2 + \Delta} \v.
	\end{align*}
	Substituted into the input (\ref{eq:penaltyController}), this indeed leads to the same result as in the case when the entire input was penalized as a whole with $\R{'}(\x) = K^{-1}(\norm[2]{\v(\x)}^2 + \Delta)\I$.
\end{proof}

\subsection{Proof of Theorem \ref{theorem:penaltyCombination}}

\begin{proof}
	Substituting in the specific form of $\R[i](\x)$ and similarly to the discussion following (\ref{eq:equivalentOpt}), the solution is equivalent to minimizing:
	\begin{align*}
	\min_{\u} \sum_{i=1}^{M} \left(\alpha_{i}\dfrac{\partial P_i(\rho^{\mu_i}(\x), t)}{\partial \rho^{\mu_i}}\v[i]\tp \u + \dfrac{1}{2}\alpha_{i}\u\tp \v[i] \v[i]\tp\u\right) \\
	+ \dfrac{1}{2}\u\tp \Delta \u,
	\end{align*}
	where we also used the identity $\sum \alpha_i = 1$ for taking out the regularization term from the summation. Setting the gradient with respect to the input to zero, this leads to the equation:
	\begin{equation*}
	\left(\sum_{i=1}^{M}\alpha_{i} \v[i] \v[i]\tp + \Delta \I\right) \u = \sum_{i=1}^{M} -\alpha_{i}\dfrac{\partial P_i(\rho^{\mu_i}(\x), t)}{\partial \rho^{\mu_i}}\v[i].
	\end{equation*}
	Again, note the resemblance of the solution
	\begin{equation*}
	\u = \left(\sum_{i=1}^{M}\alpha_{i} \v[i] \v[i]\tp + \Delta \I\right) \inv \sum_{i=1}^{M} \left(-\alpha_{i}\dfrac{\partial P_i(\rho^{\mu_i}(\x), t)}{\partial \rho^{\mu_i}}\v[i]\right)
	\end{equation*}
	to the one derived previously in (\ref{eq:improvedCombination}), as was to be shown.
\end{proof}

\subsection{Proof of Theorem \ref{theorem:unicycleController}}

The first time derivative of $P(\rho^{\mu}(\x[1]), t)$ is given as:
\begin{align*}
\dot{P}(\rho^{\mu}, t) &= \dfrac{\partial P}{\partial \rho^{\mu}}\dfrac{\partial \rho^{\mu}}{\partial \x[1]} \x<\dot>[1] + \dfrac{\partial P}{\partial t} \\
&= \dfrac{\partial P}{\partial \rho^{\mu}}\dfrac{\partial \rho^{\mu}}{\partial \x[1]} \left(\f[1](\x[1]) + \w[1] + \g[11](\x[2])\u[1]\right) + \dfrac{\partial P}{\partial t} \\
&= \dfrac{\partial P}{\partial \rho^{\mu}} \v[1](\x)\tp \u[1] + \F(\x[1], \w[1], t),
\end{align*}
where $\F(\x[1], \w[1], t)$ holds the terms independent of $\u[1]$. 

The second derivative is a complex expression, but we are only interested in the terms dependent on the second input $\u[2]$. These stem from the first term of the above derived expression, from the dependency of $\v[1](\x)$ and possibly $\u[1]$ on the entire state $\x$ and thus $\x[2]$. However, since the input $\u[1]$ is treated as a constant, this dependency only appears through $\v[1](\x)$ as:
\begin{equation}
\dfrac{\partial P(\rho^{\mu}(\x[1]), t)}{\partial \rho^{\mu}} \u[1]\tp \v<\dot>[1](\x).
\end{equation}
Keeping only the term involving $\u[2]$ when evaluating the derivative $\v<\dot>(\x)$ we have:
\begin{equation}
\dfrac{\partial P(\rho^{\mu}(\x[1]), t)}{\partial \rho^{\mu}} \u[1]\tp \dfrac{\partial \v[1](\x)}{\x[2]}g_{22}(\x)\u[2].
\end{equation}
Defining $\v[2](\x, \u[1])\tp = \u[1]\tp \frac{\partial \v[1](\x)}{\partial \x[2]} \g[22](\x)$, the original minimization problem (\ref{eq:singlePenaltyOptimizationU2}) is thus equivalent to solving:
\begin{equation}
\min_{\u[2]} \dfrac{\partial P(\rho^{\mu}(\x[1]), t)}{\partial \rho^{\mu}} \v[2](\x, \u[1])\tp\u[2] + \dfrac{1}{2}\u[2]\tp \R[2](\x)\u[2].
\end{equation}
The solution is obtained by setting the gradient with respect to $\u[2]$ to zero and takes the form:
\begin{equation}
\u[2](\x, t) = -\dfrac{\partial P(\rho^{\mu}(\x[1]), t)}{\partial \rho^{\mu}} \R[2](\x)\inv \v[2](\x, \u[1]),
\end{equation}
as was to be shown.

\section{Scenario descriptions}

We aimed at keeping as many problem-independent algorithm hyperparameters as possible the same across the examined scenarios. The \ac{STL} tasks were enforced by adding a penalty term $P^{\lambda}(\rho^{\varphi}) = \lambda \left(\rho_{\min} - \min(\rho_{\min},\rho^{\varphi})\right)^3$, where $\lambda$ is increased from $0.5$ to $5000$ throughout $K=50$ \ac{PI2} iterations using a cosine function, i.e., at the $(k)$-th iteration we have $\lambda = 0.5 + 4999.5 \cdot (1 - \cos(\pi k/K))/2$. In each iteration, $N=100$ trajectories are sampled and normalized using (\ref{eq:solution_normalization}) with parameters $\epsilon = 0.5$ and $h = 3.0$. (Note that the temperature parameter $\eta$ is eliminated when computing the weights corresponding to each cost and thus does not need to be defined.) The funnel adaptation parameters in Algorithm \ref{alg:funnelAdaptation} were set to $\xi_t = 0.8$ and $\beta = 0.2$ for the simple scenario and $\beta = 0.8$ for the complex scenario. Generally, we found that a higher $\beta$ value leads to a possibly more optimal solution to be found; however, in case of controlling the unicycle, this came at the expense of decreased algorithm stability. This is the reason why a lower $\beta$ value was used uniformly for the simple scenario.

Although the dynamical systems are different in each scenario, overlapping controller parameters were also kept the same. Thus, whether computing individual control actions according to (\ref{eq:generalPenaltyController}), (\ref{eq:unicycleU1}), or (\ref{eq:unicycleU2}), the penalty derivative is defined corresponding to (\ref{eq:functionS}) such that $-\frac{\partial P(\rho,t)}{\partial \rho} = 0.8\xi + 2.4/(1 + e^{-24(\xi - 1)})$ and the regularization parameter is set to $\Delta = 0.05$. The controllers from different robustness specifications are combined using the improved combination controller, which for each system has a form similar to (\ref{eq:improvedCombination}).

\addtolength{\textheight}{-6.7cm}

\subsection{Simple navigational task} \label{appendix:simpleScenario}

The scenario involves a robot tasked with navigating to and staying within an $r_g = 0.2$ radius goal region centered at $\x[g] = [1.0\ 3.5]\tp$ within $10$ seconds while avoiding a large circular obstacle of radius $r_o = 1.2$ centered at $\x[o] = [2.5\ 2.0]\tp$. The robot is initially located at $\x[0] = [3.5\ 0.3]\tp$ and must satisfy the task with a robustness measure of at least $\rho_{\min} = 0.05$ while minimizing given target functions. The formal \ac{STL} specification of the task is $\varphi = \phi_1 \and \phi_2 = \eventually[0][10] \always \mu_1 \and \always\mu_2$ where $\mu_1 = (r_g - \norm{\x - \x[g]} > 0)$ and $\mu_2 =  (\norm{\x - \x[o]} - r_o > 0)$. The scenario is simulated for $T = 10$s with a time step $\Delta t = 0.02$s. 

The funnels aiming to enforce the constraints are defined as follows. For avoiding the obstacle, the corresponding funnel is kept fix with $\gamma_2(t) = \rho_{\min}$ and $\Gamma_2(t) = 0.5$. For reaching the goal region, the funnel is initialized as $\gamma_1(t) = -5.0$ and $\Gamma_1(t) = 0.2$, and the former is allowed to change during funnel adaptation with a lower bound of at least ${\gamma}_1^{\lim}(t) = -7.0$.

Two cases are examined; one, where the robot is modeled as a single integrator $\x<\dot> = [\dot{x}\ \dot{y}]\tp = [u_x\ u_y]\tp = \u$ with the input constraint $\norm[2]{\u} \le 1$; and a second, where the robot is modeled as a unicycle $\dot{x} = v\cos\theta,\ \dot{y} = v\sin{\theta},\ \dot{\theta} = 5 \omega$ with input constraints $|v| \le 1$ and $|\omega| \le 1$. The exploration for the two inputs in either case is defined by initial and minimal covariances $\C[t]'(0)' = 2\I[2]\cdot10^{-4}$ and $\C[t,\min] = 2\I[2] \cdot 10^{-7}$ for all $t = 0,\dots,T$.

\subsection{Complex navigational task} \label{appendix:complexScenario}

In the complex navigation task, we consider two ground vehicles and a drone described by (noisy) single integrator dynamics and subject to the consensus protocol with additional free inputs:
\begin{equation}
\x<\dot>(t) = -0.1(\L \otimes \I[2])\x(t) + \u(t) + \w(t).
\end{equation}
The matrix $\L$ is the so-called Laplacian of the graph describing agent connections within the consensus protocol \citep{mesbahi2010graph}; assuming a complete graph it takes the form: 
\begin{equation}
\L = \bmat{2 & -1 & -1 \\ -1 & 2 & -1 \\ -1 & -1 & 2}. 
\end{equation}
These dynamics fit the system (\ref{eq:systemDynamics}) with the known input term $g(\x) = \I$ and unknown $f(\x) = -0.1(\L \otimes \I[2])$. The subscripts $\x[i]$ and $\u[i]$, $i = 1,2,3$, refer to the location and inputs of the $i$-th robot. The input constraint is $\norm[2]{\u[i]} \le 1$ for each robot. 
The robots' initial locations are $\x[1,0] = [3.0\ 0.8]\tp$, $\x[2,0] = [2.0\ 0.8]\tp$, and $\x[3,0] = [1.2\ 0.7]\tp$.

The ground robots are tasked with reaching and staying within $r_g = 0.1$ meters of $\x[g1] = [2.0 \ 4.2]\tp$ and $\x[g2] = [3.0 \ 4.2]\tp$ within $7s$ while maintaining a mutual distance between $d_{12}^{\min} = 1 - \Delta d_{12}$ and $d_{12}^{\max} =1 + \Delta d_{12}$, $\Delta d_{12} = 0.1$. Furthermore, they must avoid a circular obstacle of radius $1$m centered at $\x[o] = [2.5 \ 2.5]\tp$ by $r_o = 1.2$ during this maneuver (in order to leave space for, e.g., a carried object). The drone is tasked with reaching and staying within $r_a = 0.1$ meters from the middle of the two ground robots within 3 seconds. The goal is to satisfy this task with robustness $\rho_{\text{min}} = 0.02$ while minimizing the sum of each robot's expended energy, i.e., $C(\tau) = \sum_{i=1}^{3}\int_{0}^{T}\u[i]\tp \u[i]$. The scenario is simulated for $T = 10$s with a time step of $\Delta t = 0.01s$.  The exploration for the six inputs is defined by initial and minimal covariances $\C[t]'(0)' = 2\I[2]\cdot10^{-4}$ and $\C[t,\min] = 2\I[2] \cdot 10^{-7}$ for all $t = 0,\dots,T$.

A formal description of the task within the STL framework is given as follows. Define the non-temporal formulas $\mu_i = (\norm{\x[i] - \x[gi]} \le r_{g})$ for $i=1,2$, $\mu_3 = (\norm{\x[1] - \x[2]} \le d_{12}^{\max})$, $\mu_4 = (\norm{\x[1] - \x[2]} \ge d_{12}^{\min})$, $\mu_5 = (\norm{\x[1] - \x[o]} \ge r_{o})$, $\mu_6 = (\norm{\x[2] - \x[o]} \ge r_{o})$, and $\mu_7 = (\norm{(\x[1] + \x[2])/2 - \x[3]} \le r_{a})$. The corresponding temporal formulas are then $\phi_{i} = \eventually[0][7] \always[0][\infty] \mu_i$ for $i = 1,2$, $\phi_{i} = \always[0][\infty]\mu_i$ for $i = 3\dots 6$, and $\phi_{7} = \eventually[0][3]\always[0][\infty]\mu_7$. The full task specification is thus given as $\varphi = \bigwedge_{i=1}^{7} \phi_{i}$.

The funnels aiming to describe the evolution of the $i=1,\dots,7$ $\mu_i$ predicates are defined as follows. For reaching the goal regions, we have $\gamma_i(t) = -4.0$, $\Gamma_i(t) = 0.1$ and ${\gamma}_i^{\lim}(t) = -7.0$ for $i=1,2$. For avoiding the obstacle, we define $\gamma_i(t) = 0.0$, $\Gamma_i(t) = 0.5$ and ${\gamma}_i^{\lim}(t) = 0.0$ for $i=3,4$. For maintaining the target distance between the ground robots, we have $\gamma_i(t) = 0.0$, $\Gamma_i(t) = 0.1$ and ${\gamma}_i^{\lim}(t) = 0.0$ for $i=5,6$. Finally, for the drone we have $\gamma_7(t) = -2.0$, $\Gamma_i(t) = 0.1$ and ${\gamma}_i^{\lim}(t) = -4.0$. Note that only the funnels aiming to enforce obstacle avoidance and the distance constraints actually help towards satisfying the \ac{STL} task $\varphi$. For example, the funnels defined for reaching the goal regions does not attempt to enforce $\rho^{\mu_1} \ge 0$ within the desired $7$ seconds.

\end{document}